\documentclass[epj,nopacs]{svjour}
\usepackage[T1]{fontenc}
\usepackage{graphicx}
\usepackage[margin=2cm]{geometry}
\usepackage[affil-it]{authblk}
\usepackage[caption=false,singlelinecheck=false]{subfig}
\usepackage{float}
\usepackage{booktabs}
\usepackage{ulem}
\usepackage{amsmath}
\usepackage{amssymb}
\usepackage{amsfonts}
\usepackage{mathrsfs}
\usepackage{nicefrac}
\usepackage{newtxtext}
\usepackage{newtxmath}
\usepackage{slashed}
\usepackage{enumitem}
\usepackage{rotating}
\usepackage{cite}
\usepackage{placeins} \usepackage[colorlinks, urlcolor=black,
  citecolor=blue, link color=black]{hyperref}

\newcommand{\modulus}[1]{\ensuremath{\left| #1 \right|}}

\renewcommand{\Re}{\textrm{Re}}

\newcommand{\mrd}{\mathrm{d}}

\numberwithin{equation}{section}
\journalname{}

\def\makeheadbox{ }
\allowdisplaybreaks

\begin{document}
\title{\bf Exploring CP violation in \texorpdfstring{$\boldsymbol{H
\to \tau^+\,\tau^-\,\gamma}$}{H -> tau+ tau- gamma}}

\titlerunning{Exploring CP violation in $H
\to \tau^+\,\tau^-\,\gamma$}

\author{Erlend Aakvaag\inst{1}\thanks{Erlend.Aakvaag@uib.no}, Nikolai
Fomin\inst{1}\thanks{nikolai.fomin@cern.ch}, Anna
Lipniacka\inst{1}\thanks{anna.lipniacka@cern.ch}, Stefan
Pokorski\inst{2}\thanks{Stefan.Pokorski@fuw.edu.pl}, Janusz
Rosiek\inst{2}\thanks{Janusz.Rosiek@fuw.edu.pl} \and
Dibyakrupa Sahoo\inst{2}\thanks{Dibyakrupa.Sahoo@fuw.edu.pl
(corresponding author)}}

\authorrunning{Aakvaag, Fomin, Lipniacka, Pokorski, Rosiek and Sahoo}

\institute{Department of Physics and Technology, University of Bergen,
All\'{e}gaten 55, 5007 Bergen, Norway \and Institute of Theoretical
Physics, Faculty of Physics, University of Warsaw, ul.\ Pasteura 5,
02-093 Warsaw, Poland}

\date{\today}
\abstract{We propose a method of measuring the CP-odd part of the
Yukawa interaction of Higgs boson and $\tau$ leptons by observing the
forward-backward asymmetry in the decay $H \to \tau^+ \, \tau^- \,
\gamma$.  The source of such asymmetry is the interference of the
CP-even loop-level contribution coming from $H \to Z \, \gamma \to
\tau^+ \, \tau^- \, \gamma$ decay channel with the contribution from
tree-level CP-odd Yukawa interaction.  We find that the CP violating
effect is maximum when the invariant mass of the $\tau^+\,\tau^-$ pair
is equal to the mass of the $Z$ boson.  We propose and utilise various
Dalitz plot asymmetries to quantify the maximal size of the asymmetry
and perform Monte Carlo simulations to study the feasibility of
measuring it in the high luminosity phase of the Large Hadron Collider
(HL-LHC). }

\maketitle

\section{Introduction}\label{sec:intro}

In the Standard Model (SM), violation of the CP symmetry is encoded in
the CKM matrix.  In principle, a Beyond the Standard Model (BSM)
physics may have new sources of CP violation.  In particular, BSM CP
violation in the Yukawa interactions is welcome for electroweak
baryogenesis (it is well known that CP violation in the SM is by far
too weak for baryogenesis \cite{Shaposhnikov:1987tw, Gavela:1993ts,
Gavela:1994dt}).  The most general
expression for CP violating $H\psi\psi$ Yukawa interaction can be
written in the following form, %
\begin{equation}
\label{eq:Htautau-Lagrangian-CPV}
\mathscr{L}_{H\psi\psi}^{} = - \frac{m_\psi}{\varv} \, \overline{\psi}
\left(a_\psi + i\, \gamma^5 \, b_\psi \right) \psi \, H \,,
\end{equation}
where $\varv$ is the vacuum expectation value of the Higgs field,
$m_\psi$ denotes the mass of the fermion $\psi$, and $a_\psi, b_\psi$
are two \textit{real} valued parameters.  In the SM,
$a_\psi^\textrm{SM}=1$, $b_\psi^\textrm{SM}=0$.  If simultaneously
both $a_\psi \neq 0$ and $b_\psi \neq 0$, it implies CP violation in
$H\psi\psi$ Yukawa interaction.  The parameters $b_\psi$ are strongly
constrained by the experimental bounds on the electron and neutron
Electric Dipole Moments (for a recent analysis see
\cite{deVries:2017ncy, DeVries:2018aul, Fuchs:2020uoc, Bahl:2022yrs}
and references therein).  In this context, the $\tau$ lepton Yukawa
coupling is of interest as it is large and the EDM bound, $b_\tau
<0.3$, is weak enough for the $\tau$ Yukawa to play a role in
electroweak baryogenesis \cite{deVries:2017ncy, DeVries:2018aul,
Fuchs:2020uoc, Bahl:2022yrs}, see however
\cite{Alonso-Gonzalez:2021jsa}.  CP violation in the $\tau$ Yukawa has
also been searched for at the LHC.  The recent study by CMS
\cite{CMS:2021sdq} probing $H \to \tau^+ \, \tau^-$ gives
$\modulus{b_\tau} \lesssim 0.34$ at $68.3\%$ confidence level (for
further prospects see \cite{Gritsan:2022php}).  Majority of the
experimental studies on this issue concentrate on measurements of the
angle between $\tau$ decay planes determined by the directions of
particles produced in subsequent $\tau$ lepton decays, such as in
$H\to \tau^+ \, \tau^-\to \pi^+ \, \pi^- \, \nu_\tau \, \bar\nu_\tau$
or $H\to \tau^+ \, \tau^- \to \rho^+ \, \rho^- \, \nu_\tau \,
\bar\nu_\tau$~\cite{ATLAS:2020evk, CMS:2021sdq, CMS:2022uox,
Berge:2008wi, Berge:2008dr, Berge:2011ij, Berge:2013jra,
Harnik:2013aja, Hagiwara:2016zqz}.

\begin{figure*}[hbt!]
\centering
\subfloat[$H\tau\tau$ Yukawa contribution at tree-level.  The empty
  circles denote the fact that the final photon can arise from either
  of the $\tau$ legs.]{\hspace{15mm}\includegraphics[scale=0.8]{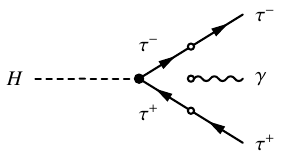}\hspace{15mm}} \hfil
\subfloat[$Z\gamma$ and $\gamma\gamma$ contributions from triangular
  top loop.]{\hspace{15mm}\includegraphics[scale=0.8]{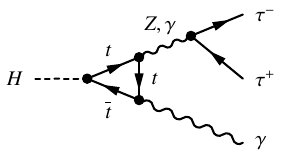}\hspace{15mm}} \\
\subfloat[$Z\gamma$ and $\gamma\gamma$ contribution from triangular
  $W$ loop.]{\hspace{15mm}\includegraphics[scale=0.8]{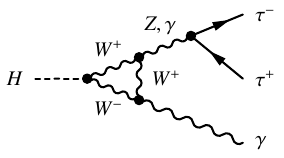}\hspace{15mm}}
\hfil
\subfloat[$Z\gamma$ and $\gamma\gamma$ contribution from $W$ loop with
  $WW\gamma Z/\gamma$
  vertex.]{\hspace{15mm}\includegraphics[scale=0.8]{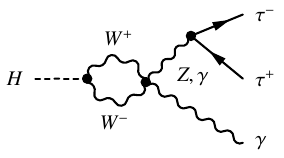}\hspace{15mm}}
\caption{The dominant Feynman diagrams contributing to $H \to \tau^+
  \, \tau^- \, \gamma$.  The contributions of possible box diagrams at
  one-loop level (not shown here) are negligible, see
  Refs.~\cite{Abbasabadi:1995rc, Abbasabadi:1996ze}.}
\label{fig:FD-H2ttg}
\end{figure*}

In this paper we propose to measure the for\-ward-back\-ward asymmetry of
$\tau$ lepton angular distribution in the $H \to \tau^+ \, \tau^- \,
\gamma$ decay, as a measure of the CP violation in $H\tau\tau$ Yukawa
interaction.  To study this asymmetry we utilise the Lorentz invariant
Dalitz plot distribution of events.  The dominant CP-violating effects
which contribute to the for\-ward-back\-ward asymmetry in the $H \to
\tau^+ \, \tau^- \, \gamma$ decay are proportional to the interference
of the tree-level and loop-level diagrams\footnote{One could, in
principle, also consider $H \to \ell^+ \, \ell^- \, \gamma$, with
$\ell=e,\mu$, \cite{Chen:2014ona, Korchin:2014kha}, facilitated by
similar Feynman diagrams as in Fig.~\ref{fig:FD-H2ttg}, to probe CP
violation in the corresponding Yukawa interactions.  But for these
processes the loop-level contributions are dominant and can overshadow
the CP-violating part of the tiny Yukawa interaction of $e,\mu$ with
Higgs boson.  Nevertheless, our proposal to use the Lorentz invariant
Dalitz plot distribution to study forward-backward asymmetry holds for
these decays as well.  In such a case, observation of a sizeable
asymmetry would suggest CP violation in the loop-level contributions.}
shown in Fig.~\ref{fig:FD-H2ttg}.  The lower branching ratio than the
$H \to \tau^+ \, \tau^-$ is partially compensated by the fact that one
only requires to reconstruct the 4-momenta of the $\tau$ leptons and
not the full spatial distributions of the final $\tau$ decay products.
Our heuristic simulations for the HL-LHC show that one can possibly
probe $b_\tau$ using our proposed methodology.  A more thorough Monte
Carlo study scanning the full 2-dimensional Dalitz plot distribution
is beyond our current expertise, and is hence reserved for future
exploration.

Our paper is organised as follows.  In Sec.~\ref{sec:pheno} we briefly
outline the important phenomenological aspects of the 3-body decay $H
\to \tau^+ \, \tau^- \, \gamma$, showing how the for\-ward-back\-ward
asymmetry originates and how can it be probed from the Lorentz
invariant Dalitz plot distribution.  In Sec.~\ref{sec:numerical-study}
we do a numerical study, looking at the distribution pattern inside
the Dalitz plot and assess how large the forward-backward asymmetry
could be.  In Sec.~\ref{sec:simulation} we perform a heuristic Monte
Carlo study of the feasibility of observing the asymmetry in context
of HL-LHC.  Finally we conclude in Sec.~\ref{sec:conclusion}
summarising our findings and highlighting the salient features of our
proposed methodology.

\section{Phenomenological study of \texorpdfstring{$\boldsymbol{H \to
\tau^+ \, \tau^- \, \gamma}$}{H -> tau+ tau- gamma}}\label{sec:pheno}

\begin{figure*}[htbp]
\centering%
\includegraphics[scale=0.8]{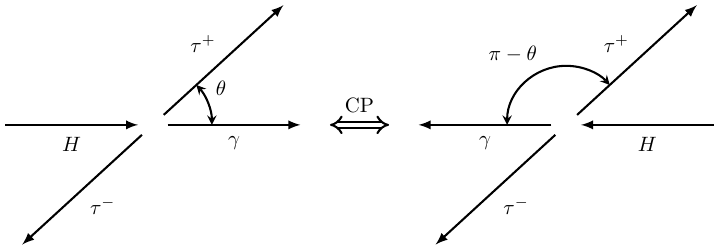}
\caption{Kinematic configurations related by CP in the
  center-of-momentum frame of $\tau^+ \, \tau^-$.}
\label{fig:CoM-kinematics-CP}
\end{figure*}

The decay $H \to \tau^+ \, \tau^- \, \gamma$ is its own CP-conjugate
process.  Let us study the kinematic configuration of the decay in the
center-of-momentum frame of $\tau^+ \, \tau^-$ (equivalently called
the di-tau rest frame).  From Fig.~\ref{fig:CoM-kinematics-CP} it is
clear that the CP transformation takes the angle $\theta$ between
$\tau^+$ and photon to $\pi-\theta$.  This implies that any difference
(or asymmetry) in the angular distribution of events with respect to
$\cos\theta \leftrightarrow -\cos\theta$ (`forward' $\leftrightarrow$
`backward') exchange would be a clear signature of CP-violation.

As illustrated in Fig.~\ref{fig:FD-H2ttg} the decay $H \to \tau^+ \,
\tau^- \, \gamma$ proceeds via the tree-level $H\tau\tau$ Yukawa
interaction, as well as via the effective vertex of $H \to
\mathcal{V}\,\gamma \to \tau^+ \, \tau^- \, \gamma$, with
$\mathcal{V}=Z,\gamma$.  The effective Lagrangian for the later
interaction can be, to the lowest mass dimension order, written in the
form,
\begin{align}
\mathscr{L}_{H\mathcal{V}\gamma} &= \frac{H}{4\,\varv} \Big(2
A_2^{Z\gamma} F^{\mu\nu} Z_{\mu\nu} + 2 A_3^{Z\gamma} F^{\mu\nu}
\widetilde{Z}_{\mu\nu} \nonumber\\*%
& \qquad + A_2^{\gamma\gamma} F^{\mu\nu} F_{\mu\nu} +
A_3^{\gamma\gamma} F^{\mu\nu} \widetilde{F}_{\mu\nu} \Big),
\label{eq:HVgamma-Lagrangian-CPV}
\end{align}
where $\mathcal{V}_{\mu\nu} = \partial_\mu \mathcal{V}_\nu -
\partial_\nu \mathcal{V}_\mu$, $\widetilde{\mathcal{V}}_{\mu\nu} =
\frac{1}{2} \epsilon_{\mu\nu\rho\sigma} \mathcal{V}^{\rho\sigma}$, and
$A_{2,3}^{\mathcal{V}\gamma}$ are dimensionless form factors.  Such
form factors receive contributions from the SM loop-level diagrams
(see Fig.~\ref{fig:FD-H2ttg}), and from the interaction beyond the SM,
the latter in general possibly also containing CP-violating couplings.
We take into account only the SM loop contributions, assuming that BSM
loop corrections are small compared to the tree level ones.  Thus, we
put $A_3^{\mathcal{V}\gamma}=0$ while doing numerical
study\footnote{Note that for top quark or $W$ boson contributions to
$H\mathcal{V}\gamma$ coupling, loop integrals are purely real, so the
CP violating form factors can only be proportional to imaginary
couplings.}, but for completeness we will keep the
$A_3^{\mathcal{V}\gamma}$ dependent terms in our analytical
expressions.  The expressions for $A_2^{Z\gamma}$ and
$A_2^{\gamma\gamma}$ in the SM are given in
Ref.~\cite{Bergstrom:1985hp}.

Let us denote the decay amplitude for $H\to\tau^+\,\tau^-\,\gamma$ by
$\mathscr{M}$.  As illustrated in Fig.~\ref{fig:FD-H2ttg}, the
amplitude can be split into three parts: (1) tree-level contribution
$\mathscr{M}^{\textrm{(Yuk)}}$, (2) loop-level $Z\gamma$ contribution
$\mathscr{M}^{(Z\gamma)}$, and (3) loop-level $\gamma\gamma$
contribution $\mathscr{M}^{(\gamma\gamma)}$, i.e.\ $\mathscr{M} =
\mathscr{M}^{\textrm{(Yuk)}} + \mathscr{M}^{(Z\gamma)} +
\mathscr{M}^{(\gamma\gamma)}$.  Like any other 3-body decay of a
spin-0 particle, the full kinematics of $H(p_H) \to \tau^+ (p_+) \,
\tau^- (p_-) \, \gamma (p_0)$ can be described by two independent
variables.  We choose to work with Lorentz invariant mass squares.
Defining
\begin{subequations}
\label{eq:InvMassSq}
\begin{align}
m_{+-}^2 &= \left(p_+ + p_-\right)^2 = \left(p_H - p_0\right)^2,
\label{eq:mpmSq}\\ %
m_{+0}^2 &= \left(p_+ + p_0\right)^2 = \left(p_H - p_-\right)^2,
\label{eq:mpzSq}\\ %
m_{-0}^2 &= \left(p_- + p_0\right)^2 = \left(p_H -p_+\right)^2,
\label{eq:mmzSq}
\end{align}
\end{subequations}
where
\begin{equation}\label{eq:mpz2mmz2mpm2}
m_{+-}^2 + m_{+0}^2 + m_{-0}^2 = m_H^2 + 2 \, m_\tau^2.
\end{equation}
We can express $\cos\theta$, defined in the di-tau rest frame, in
terms of the Lorentz invariant variables:
\begin{equation}\label{eq:costheta}
\cos\theta = \left( 1 - \dfrac{4\,m_\tau^2}{m_{+-}^2}
\right)^{-\frac{1}{2}} \, \frac{m_{-0}^2 - m_{+0}^2}{m_H^2 -
m_{+-}^2}\, .
\end{equation}

At the beginning of this section we have argued that the
forward-backward asymmetry in $\cos\theta$ distribution can serve as a
probe of CP violation.  Therefore, we see that the forward-backward
asymmetry would be equivalent to an asymmetry in the distribution or
number of events in the $m_{+0}^2$ vs.\ $m_{-0}^2$ plane (usually
called a Dalitz plot) under the exchange $m_{+0}^2 \leftrightarrow
m_{-0}^2$.  Equivalently, one can consider distribution of events in
the $m_{+0}$ vs.\ $m_{-0}$ plane which may be more convenient from
experimental perspective.  The `forward' (or `backward') region in
Dalitz plot is that region where $m_{-0} > m_{+0}$ (or $m_{-0} <
m_{+0}$).

In the rest frame of the Higgs boson, the differential decay rate of
$H \to \tau^+\,\tau^-\,\gamma$ in terms of $m_{+0}$ and $m_{-0}$ is
given by,
\begin{equation}
\label{eq:diff-decay-rate-1}
\dfrac{\mrd^2 \Gamma_{\tau\tau\gamma}}{\mrd m_{+0} \, \mrd m_{-0}} =
\frac{m_{+0} \, m_{-0}}{64 \, \pi^3 \, m_H^3} \modulus{\mathscr{M}}^2
\equiv \mathcal{D} \big(m_{+0},m_{-0}\big),
\end{equation}
where the squared amplitude $\modulus{\mathscr{M}}^2$ can be split
into six constituents,
\begin{align}
\modulus{\mathscr{M}}^2 &= \big|\mathscr{M}^{(\textrm{Yuk})}\big|^2 +
\big|\mathscr{M}^{(Z\gamma)}\big|^2 +
\big|\mathscr{M}^{(\gamma\gamma)}\big|^2 \nonumber\\%
&\quad + 2 \, \Re\left(\mathscr{M}^{(\gamma\gamma)} \,
\mathscr{M}^{(Z\gamma)*}\right) \nonumber\\%
&\quad + 2 \, \Re\left(\mathscr{M}^{(\textrm{Yuk})} \,
\mathscr{M}^{(Z\gamma)*}\right) \nonumber\\%
&\quad + 2 \, \Re\left(\mathscr{M}^{(\textrm{Yuk})} \,
\mathscr{M}^{(\gamma\gamma)*}\right).\label{eq:ampSq-decomposition}
\end{align}
In order to clearly point out the terms responsible for the
forward-backward asymmetry and see how it is related to CP-asymmetry,
we write down the expression for the individual constituents of
amplitude square, as shown in Eq.~\eqref{eq:ampSq-decomposition}, in
terms of $m_{+-}^2$ and $\theta$.  Using Eqs.~\eqref{eq:mpz2mmz2mpm2}
and \eqref{eq:costheta} one can easily rewrite all these expressions
in terms of $m_{+0}$ and $m_{-0}$.  Neglecting the subdominant
$m_\tau$ dependent terms in the numerator, we have:
\begin{subequations}\label{eq:individual-ampSq-components}
\begin{align}
&\big|\mathscr{M}^{(\textrm{Yuk})}\big|^2 = \frac{16 e^2
\left(a_{\tau}^2+b_{\tau}^2\right) m_{\tau}^2
\left(m_H^4+m_{+-}^4\right) \, m_{+-}^4 \, \sin^2\theta}{\varv^2
\left(m_H^2-m_{+-}^2\right)^2 \left( \left(m_{+-}^2 -
4\,m_\tau^2\right) \, \sin^2\theta + 4\,m_\tau^2 \right)^2},\\
&\big|\mathscr{M}^{(Z\gamma)}\big|^2 = \frac{g_Z^2 \, \left(\left(
c_A^\tau \right)^2+\left( c_V^\tau \right)^2\right) \, \left(\left(
A_2^{Z\gamma} \right)^2+\left( A_3^{Z\gamma} \right)^2\right)}{8
\varv^2 \left(\left(m_{+-}^2-m_Z^2\right)^2+\Gamma_Z^2 m_Z^2\right)}
\nonumber\\%
&\hspace{15mm} \times m_{+-}^2 \left(m_H^2-m_{+-}^2\right)^2
\left(1+\cos^2\theta\right),\\
&\big|\mathscr{M}^{(\gamma\gamma)}\big|^2 = \frac{e^2 \left(\left(
A_2^{\gamma\gamma} \right)^2+\left( A_3^{\gamma\gamma}
\right)^2\right)}{2 \, m_{+-}^2 \, \varv^2} \nonumber\\%
&\hspace{15mm} \times \left(m_H^2-m_{+-}^2\right)^2
\left(1+\cos^2\theta\right),\\ %
&\Re\left(\mathscr{M}^{(\gamma\gamma)} \,
\mathscr{M}^{(Z\gamma)*}\right) = -\frac{e \, g_Z \,
\left(m_H^2-m_{+-}^2\right)^2}{4 \varv^2
\left(\left(m_{+-}^2-m_Z^2\right)^2+\Gamma_Z^2 m_Z^2\right)}
\nonumber\\* %
&\quad \times \Bigg(2 \, c_A^\tau \, \left( A_2^{\gamma\gamma}
A_3^{Z\gamma}-A_2^{Z\gamma} A_3^{\gamma\gamma} \right) \, m_Z \,
\Gamma_Z \, \cos \theta \, \nonumber\\* %
&\qquad + c_V^\tau \, \left(A_2^{\gamma\gamma}
A_2^{Z\gamma}+A_3^{\gamma\gamma} A_3^{Z\gamma}\right) \,
\left(m_{+-}^2-m_Z^2\right) \left(1+\cos^2 \theta \right)\Bigg),\\%
&\Re\left(\mathscr{M}^{(\textrm{Yuk})} \,
\mathscr{M}^{(Z\gamma)*}\right) = \frac{4 \, e \, g_Z \, m_{\tau}^2 \,
m_{+-}^4 \, \sin^2\theta}{\varv^2
\left(\left(m_{+-}^2-m_Z^2\right)^2+\Gamma_Z^2 m_Z^2\right)} \nonumber\\* %
&\quad \times \frac{1}{\left(
\left( m_{+-}^2 - 4\,m_\tau^2 \right) \, \sin^2\theta + 4\,m_\tau^2
\right)^2} \nonumber\\*%
&\quad \times \Bigg( c_A^\tau \, \left( A_3^{Z\gamma} a_{\tau} -
A_2^{Z\gamma} b_{\tau} \right) \, m_Z \, \Gamma_Z \,
\left(m_H^2-m_{+-}^2\right) \, \cos\theta \nonumber\\* %
&\qquad + c_V^\tau \left(m_{+-}^2-m_Z^2\right) \bigg(A_2^{Z\gamma}
a_{\tau} \left(m_H^2-m_{+-}^2 \cos^2 \theta \right) \nonumber\\* %
&\hspace{35mm} + A_3^{Z\gamma} b_{\tau} \left(m_H^2-m_{+-}^2\right)
\bigg)\Bigg),\label{eq:IntYukZGa}\\%
&\Re\left(\mathscr{M}^{(\textrm{Yuk})} \,
\mathscr{M}^{(\gamma\gamma)*}\right) = - \frac{8 \, e^2 \, m_{\tau}^2
\, m_{+-}^2 \, \sin^2\theta}{\varv^2 \, \left( \left(m_{+-}^2 - 4\,
m_\tau^2 \right) \, \sin^2\theta + 4\, m_\tau^2 \right)^2}
\nonumber\\* %
&\quad \times \left(A_2^{\gamma\gamma} a_{\tau} \left(m_H^2 - m_{+-}^2
\, \cos^2\theta\right) + A_3^{\gamma\gamma} b_{\tau}
\left(m_H^2-m_{+-}^2\right) \right),
\end{align}
\end{subequations}
where $c_V^\tau = -1/2 + 2 \, \sin^2\theta_W$, $c_A^\tau = -1/2$, and
$g_Z = e/(\sin\theta_W \, \cos\theta_W)$, with $\theta_W$ being the
weak mixing angle.  Note that we have kept the total width of the $Z$
boson, $\Gamma_Z$, because the $Z$ boson can be on-shell in our case.

We are interested in terms that are odd (linear) in $\cos\theta$ (or,
using Lorentz invariant variables, odd in the difference $m_{+0}^2 -
m_{-0}^2$).  Such terms are found to be proportional to $m_Z \,
\Gamma_Z$ as well as the product of CP-even and CP-odd couplings.  If
we use the narrow-width approximation for the $Z$ boson propagator,
\begin{equation}
\label{eq:Narrow-Width-Approximation}
\frac{1}{\left(m_{+-}^2-m_Z^2\right)^2 + \Gamma_Z^2 \, m_Z^2} \approx
\frac{\pi}{m_Z \, \Gamma_Z} \delta\big(m_{+-}^2-m_Z^2\big),
\end{equation}
the $m_Z \, \Gamma_Z$ factor in the terms linear in $\cos\theta$
cancels out, and it is obvious that maximum CP-violation occurs for
$m_{+-}^2 = m_Z^2$.  Thus, the dominant contribution to the
forward-backward asymmetry comes from the events for which invariant
mass of the $\tau$ pair is close to the $Z$ boson mass.

It is clear from Eq.~\eqref{eq:IntYukZGa} that to a good approximation
the asymmetry in the $\cos\theta$ distribution probes the combination
$(A_3^{Z\gamma}a_\tau-A_2^{Z\gamma}b_\tau)$.  In our numerical study
in Sec.~\ref{sec:numerical-study} we put $A_3^{Z\gamma}=0$.

In the following section we illustrate how the distribution pattern in
the `forward' and `backward' regions of the Dalitz plot differ due to
CP violation (i.e.\ $b_\tau \neq 0$) by studying the following
distribution asymmetry,
\begin{equation}
\label{eq:diff-asym}
\mathcal{A}\left(m_{+0},m_{-0}\right) =
\frac{\big|\mathcal{D}\left(m_{+0},m_{-0}\right) -
  \mathcal{D}\left(m_{-0},m_{+0}
  \right)\big|}{\mathcal{D}\left(m_{+0},m_{-0}\right) +
  \mathcal{D}\left(m_{-0},m_{+0}\right)}.%
\end{equation}
Additionally, we also study the asymmetry integrated over the region
where the invariant mass of the $\tau^+\,\tau^-$ pair is close to the
$Z$ boson mass,
\begin{equation}\label{eq:Z-pole-asym}
A(n) = \frac{\displaystyle \Big| \iint \Big(\mathcal{D}\big(m_{+0}<
  m_{-0}\big) - \mathcal{D}\big(m_{+0}>m_{-0} \big) \Big) \,
  \Pi(m_{+-},n) \, \mrd m_{+0} \, \mrd m_{-0} \Big|}{\displaystyle \iint
  \mathcal{D}\big(m_{+0},m_{-0}\big) \, \Pi(m_{+-},n) \, \mrd m_{+0} \,
  \mrd m_{-0}}, 
\end{equation}
where the function $\Pi\left(m_{+-}, n\right)$ defines the cut on the
invariant mass of the $\tau^+\,\tau^-$ pair
\begin{equation}\label{eq:Z-pole-cut}
\Pi\left(m_{+-}, n\right) =
\begin{cases}
1 & \textrm{for } \big| m_{+-} - m_Z \big| \,  \leqslant  n \, \Gamma_Z ,\\%
0 & \textrm{otherwise}.
\end{cases}
\end{equation}
The asymmetry $A(n)$ is directly related to the number of events
around the $Z$ pole,
\begin{equation}\label{eq:an}
  A(n) = \frac{\modulus{N_F(n)-N_B(n)}}{N_F(n) + N_B(n)},
\end{equation}
where $N_{F/B}(n)$ denote the number of events contained in the
forward/backward region which are also contained in the region around
$Z$ pole as defined in Eq.~\eqref{eq:Z-pole-cut}.

\section{Numerical study}\label{sec:numerical-study}

In this section we do a numerical study of the effect of the CP
violating parameter $b_\tau$ on the Dalitz plot distribution in
$m_{+0}$ vs.\ $m_{-0}$ plane.  Especially we focus on the size of the
asymmetries $\mathcal{A}\left(m_{+0},m_{-0}\right)$ and $A(n)$ as
defined in Eqs.~\eqref{eq:diff-asym} and \eqref{eq:Z-pole-asym}.  As
detailed below, we impose a few kinematic cuts in the Higgs rest
frame.  In the next section we present the results of a heuristic
simple MC simulation as an attempt to be closer to the experimental
conditions at the HL-LHC.

We note that by neglecting $m_\tau$ in comparison with Higgs mass
$m_H$, one can constrain (to a very good approximation) the sum
$a_\tau^2 + b_\tau^2$ from the experimentally measured $pp \to H \to
\tau^+ \, \tau^-$ cross-section~\cite{ATLAS:2022yrq}, which yields
\begin{equation}\label{eq:atau2_Plus_btau2}
a_\tau^2 + b_\tau^2 \approx 0.93^{+0.14}_{-0.12},
\end{equation}
where the experimental errors have been added in quadrature.

\begin{figure}[hbtp]
\centering%
\includegraphics[scale=0.8]{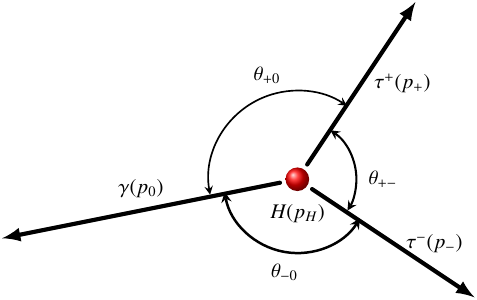}%
\caption{A schematic kinematic configuration of $\tau^+ \, \tau^- \,
  \gamma$ in the Higgs rest frame showing angles subtended by the
  3-momenta of the final particles.}%
\label{fig:H-rest-frame}%
\end{figure}

To avoid infrared divergence, we impose a cut on the photon energy
(i.e.\ specify a minimum energy for the photon) in the Higgs rest
frame,
\begin{equation}\label{eq:photon-energy-cut} 
E_\gamma^\textrm{cut} = 5~\textrm{GeV}.
\end{equation}
As we discuss later, the actual value of this cut has little impact on
the decay branching ratios in the range of the di-$\tau$ invariant
mass squared $m_{+-}^2$ most sensitive to the CP violation effect.
For the sake of reference we note that, with this cut, for the full
kinematical range of $m^2_{+-}$ the branching ratio of $H \to \tau^+
\, \tau^- \, \gamma$ is $BR_{\tau\tau\gamma}=3.72\times 10^{-3}$.  The
branching ratio decreases once a cut is imposed on the three relative
angles $\theta_X$, with $X \in \{ +-,+0,-0 \}$ (see
Fig.~\ref{fig:H-rest-frame}) among the final particles in the Higgs
rest frame.  An angular cut $\theta_X^{\textrm{cut}}$ specifies the
minimum angle among the final particles.  For $\theta_X^{\textrm{cut}}
= 5^\circ$ we get $BR_{\tau\tau\gamma}=3.24\times 10^{-3}$ which
further decreases by approximately $15$\% for each $5^\circ$ increase
in the cut.  Both the angular cut $\theta_X^{\textrm{cut}}$ and photon
energy cut $E_\gamma^\textrm{cut}$ affect the allowed values of
$m_{+0}$ and $m_{-0}$.

\begin{figure*}[hbtp]
\centering%
\begin{tabular}{cc}
\includegraphics[width=0.495\linewidth]{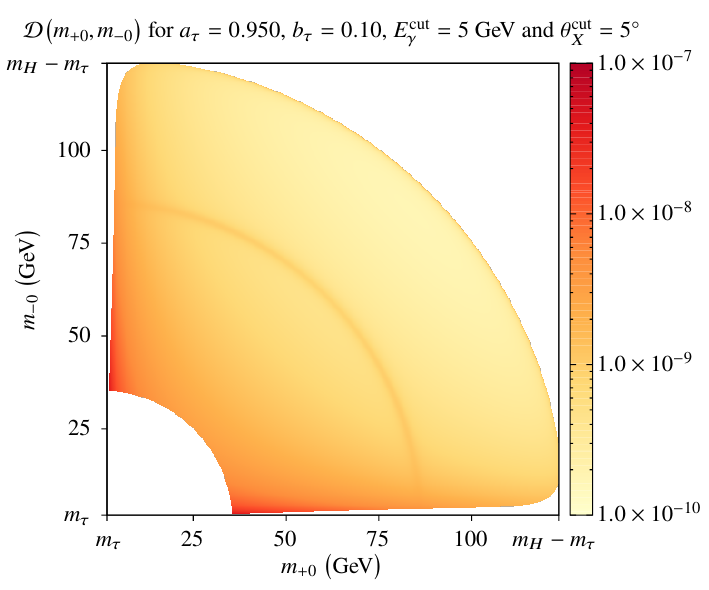} &
\includegraphics[width=0.495\linewidth]{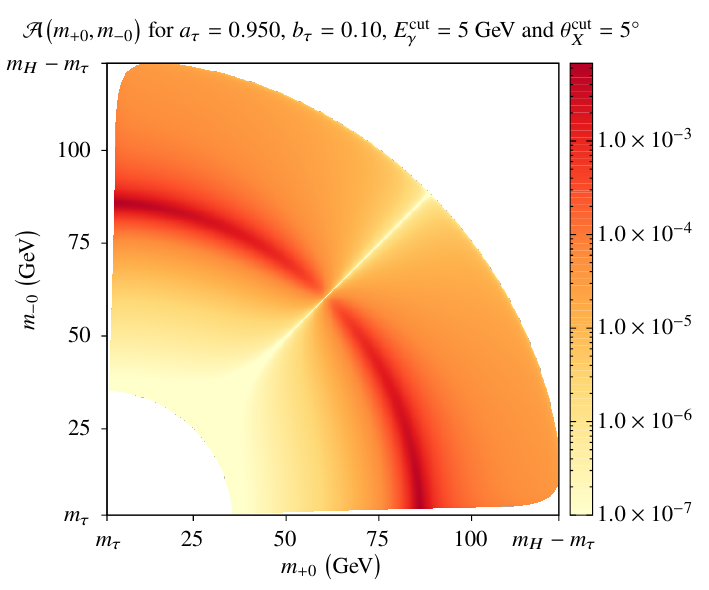}\\
\end{tabular}%
\caption{The expected differential decay rates $\mathcal{D}\big(
  m_{+0}, m_{-0} \big)$ as well as the asymmetry $\mathcal{A} \big(
  m_{+0}, m_{-0} \big)$ inside the Dalitz plot regions for $a_\tau
  =0.950$, $b_\tau=0.1$.  The asymmetry is localised around the $Z$
  pole as expected (see discussion surrounding
  Eq.~\eqref{eq:Narrow-Width-Approximation}).  Note that the colour bar
  is in logarithmic scale.}%
\label{fig:NP1-dist-asym}
\end{figure*}

In Fig.~\ref{fig:NP1-dist-asym} we see that the differential decay
distribution have maxima close to the axes when $m_{\pm 0}^2$
approaches $m_\tau^2$.  These peaks are characteristic of the
tree-level contribution from Fig.~\ref{fig:FD-H2ttg}.  A second peak
is also easily discernible in the distributions around $m_{+-}^2 =
m_Z^2$ as a slightly darker band, and this corresponds to contribution
from the on-shell $Z$ contribution, coming from the one-loop level
diagrams of Fig.~\ref{fig:FD-H2ttg}.  Furthermore, for $b_\tau \neq 0$
we do find non-zero forward-backward asymmetry.  Also as expected, the
distribution asymmetry $\mathcal{A}\left(m_{+0},m_{-0}\right)$ become
significantly large around the $Z$-pole region.  The distribution
asymmetry can be as large as $\sim 1\%$ depending on the values of
$a_\tau$, $b_\tau$ such as for $a_\tau=0.950$ and $b_\tau=0.20$.

Regarding the asymmetries $A(n)$ around the $Z$-pole, see
Eq.~\eqref{eq:Z-pole-asym}, we note that the $Z$-pole cut as encoded
in Eq.~\eqref{eq:Z-pole-cut} can be rewritten, in terms of the photon
energy in the Higgs rest frame, as follows,
\begin{align}
\Pi \big( m_{+-},n \big) &\equiv \Pi \big( E_\gamma,n \big)
\nonumber\\%
&=
\begin{cases}
1 & \textrm{ for } \Big\lvert \sqrt{m_H^2 - 2\,m_H\,E_\gamma} - m_Z
\Big\rvert \leqslant n\,\Gamma_Z,\\ %
0 & \textrm{ otherwise},
\end{cases} \label{eq:Z-pole-photon-energy}
\end{align}
From the equation above it is clear that for the invariant mass of the
$\tau$ pair close to the $Z$ pole, say $\big| m_{+-} - m_Z \big| \,
\leqslant 5 \, \Gamma_Z$, that the photon energy cut
$E_\gamma^{\textrm{cut}}=5$~GeV has no relevance, since the minimum
photon energy required for events around $Z$-pole corresponds to
higher photon energies.  Only the angular cuts
$\theta_X^{\textrm{cut}}$ have any bearing in such a case.

\begin{figure}[hbtp]
\centering%
\includegraphics[width=0.9\linewidth,keepaspectratio]{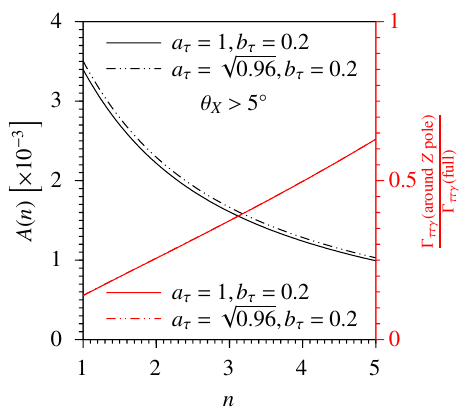}%
\caption{The comparison of forward-backward asymmetry $A(n)$, defined
  in Eq.~\eqref{eq:an} and the ratio $\Gamma_{\tau\tau\gamma}
  \textrm{(around $Z$ pole)}/\Gamma_{\tau\tau\gamma} \textrm{(full)}$,
  where $\Gamma_{\tau\tau\gamma} \textrm{(around $Z$ pole)}$ is the
  partial decay rates for the decay $H \to \tau^+ \, \tau^- \, \gamma$
  with $m_{+-}$ around the $Z$ pole satisfying
  Eq.~\eqref{eq:Z-pole-cut}, and $\Gamma_{\tau\tau\gamma}
  \textrm{(full)}$ is the full partial decay rate.}%
\label{fig:Z-pole_AFB_vs_Gttg}%
\end{figure}

In Fig.~\ref{fig:Z-pole_AFB_vs_Gttg} we show the variation of $A(n)$
for $1 \leqslant n \leqslant 5$ and compare it with with the ratio
$\Gamma_{\tau\tau\gamma} \textrm{(around $Z$
  pole)}/\Gamma_{\tau\tau\gamma} \textrm{(full)}$, where
$\Gamma_{\tau\tau\gamma} \textrm{(around $Z$ pole)}$ is the partial
decay rates for the decay $H \to \tau^+ \, \tau^- \, \gamma$ with
$m_{+-}$ around the $Z$ pole (imposed using
Eq.~\eqref{eq:Z-pole-cut}), and $\Gamma_{\tau\tau\gamma}
\textrm{(full)}$ is the full partial decay rate.  As expected, the
asymmetry decreases with $n$, as it is strongly localised around the
$Z$-pole, whereas $\Gamma_{\tau\tau\gamma} \textrm{(around $Z$
  pole)}/\Gamma_{\tau\tau\gamma} \textrm{(full)}$ increases with $n$.
The plot clearly shows the challenge for an experimental analysis to
find an optimal balance between the magnitude of the effect and the
statistics of the events.


\section{Simulation study of \texorpdfstring{$\boldsymbol{H \to \tau^+
\, \tau^- \, \gamma}$}{H -> tau+ tau- gamma} in the context of HL-LHC}
\label{sec:simulation}

To estimate the sensitivity of the proposed Higgs boson decay $H \to
\tau^+ \, \tau^- \, \gamma$ to the CP violation effects at the HL-LHC,
Monte-Carlo (MC) generators were used to simulate the signal in the
actual experimental environment.  However, due to the limited
computing resources, we have used a simplified MC simulation
procedure.  The differential cross-sections corresponding to the
various $a_\tau$ and $b_\tau$ values are computed using \textsf{GNU
  Octave}~\cite{GNUOctave} as a function of $m_{+0}$ and $m_{-0}$.
The MC signal samples are re-weighted using these cross-sections
(which include the kinematic cuts of Sec.~\ref{sec:numerical-study})
to properly model the impact of the interference term, similar to the
"interpolation" approach used in ~\cite{ATLAS:2021jbf}.  The validity
of the approach is verified by the comparison of relevant kinematic
distributions with the analytical calculations.

We project the Dalitz plot distribution of events in forward and
backward regions onto the $m_{+-}$ axis to do a 1-dimensional binned
study of the forward-backward asymmetry.  A more detailed and thorough
MC study taking the full 2-dimensional Dalitz plot distribution into
account and exploring unbinned Dalitz plot analysis techniques such as
the Miranda method \cite{Bediaga:2009tr,BaBar:2008xzl}, the method of
energy test statistic \cite{aslan2005new, Williams:2011cd,
  LHCb:2014nnj, LHCb:2023mwc, LHCb:2023rae} and the earth mover's
distance~\cite{Davis:2023lxq} are reserved for future explorations.

In the following, all additional cuts are defined in the laboratory
frame.  Reconstruction of the Higgs rest frame, that was used in the
previous section would require the knowledge of the Higgs boson
three-momentum which is not known experimentally.  Besides, the
observed distribution of events in $m_{+0}$ vs.\ $m_{-0}$ Dalitz plot
can be obtained in any frame of reference.

\subsection{Monte Carlo Simulation}\label{sec:mc}

\begin{figure*}
\centering
\subfloat[Invariant (true) di-$\tau$
  mass.]{\includegraphics[width=0.45\linewidth]{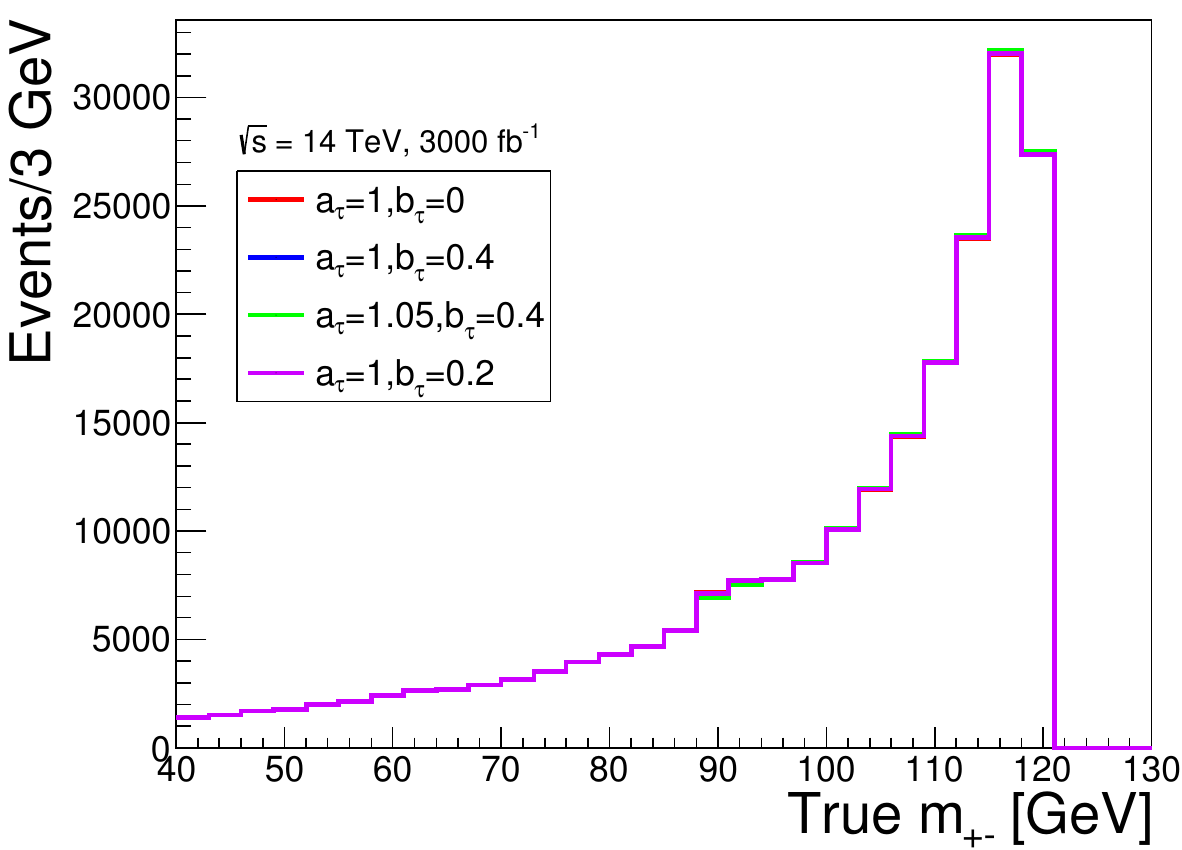}\label{fig:massplot1}} \hspace{5mm} 
\subfloat[Invariant (true) di-$\tau$ mass after kinematic
  cuts.]{\includegraphics[width=0.45\linewidth]{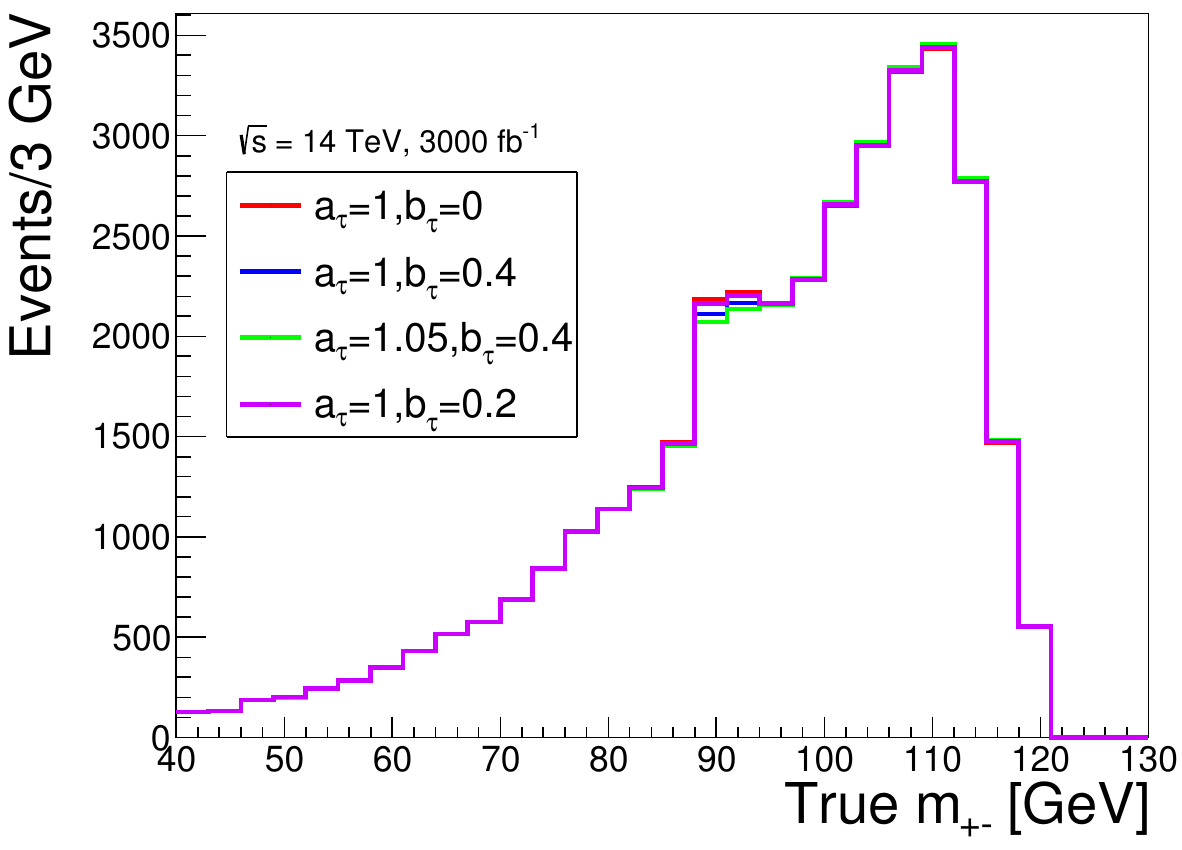}\label{fig:massplot2}}\\
\subfloat[True asymmetry after kinematic
  cuts.]{\includegraphics[width=0.45\linewidth]{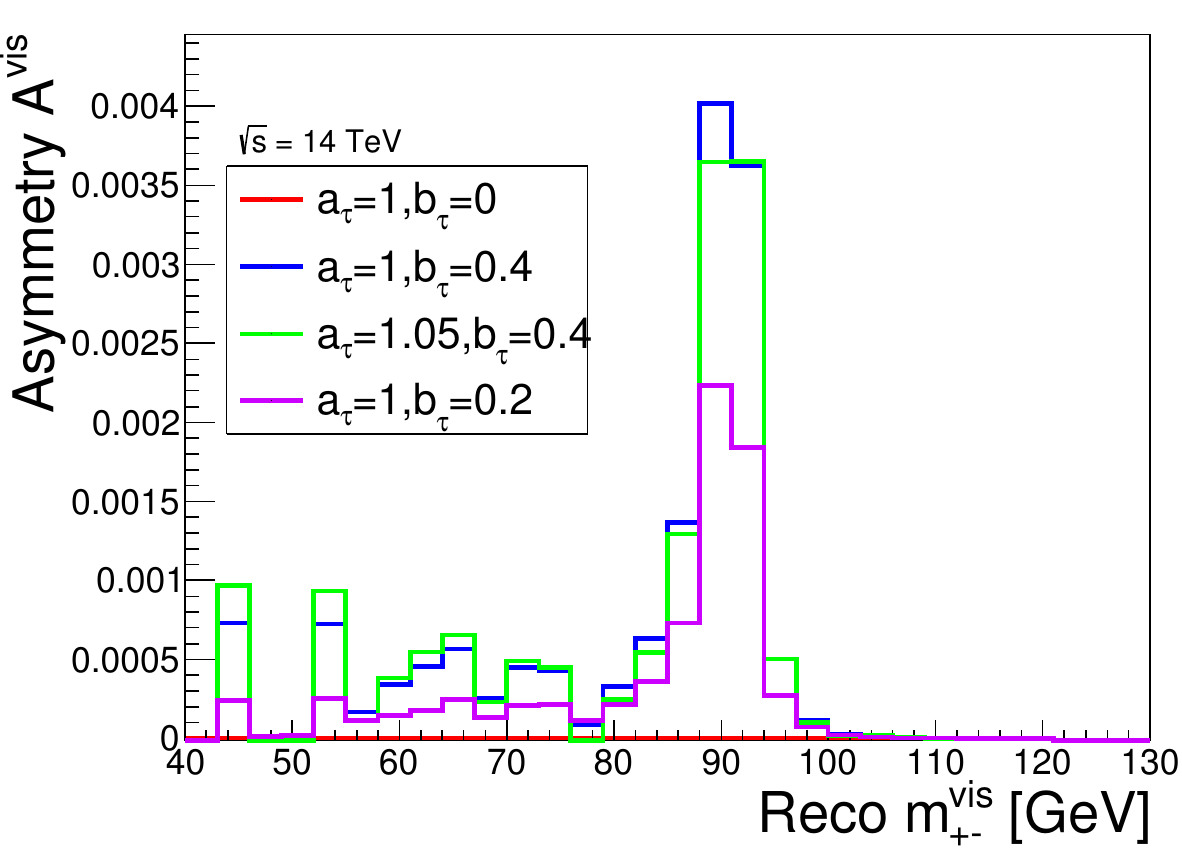}\label{fig:recoplot1}} \hspace{5mm} 
\subfloat[Visible asymmetry after kinematic
  cuts.]{\includegraphics[width=0.45\linewidth]{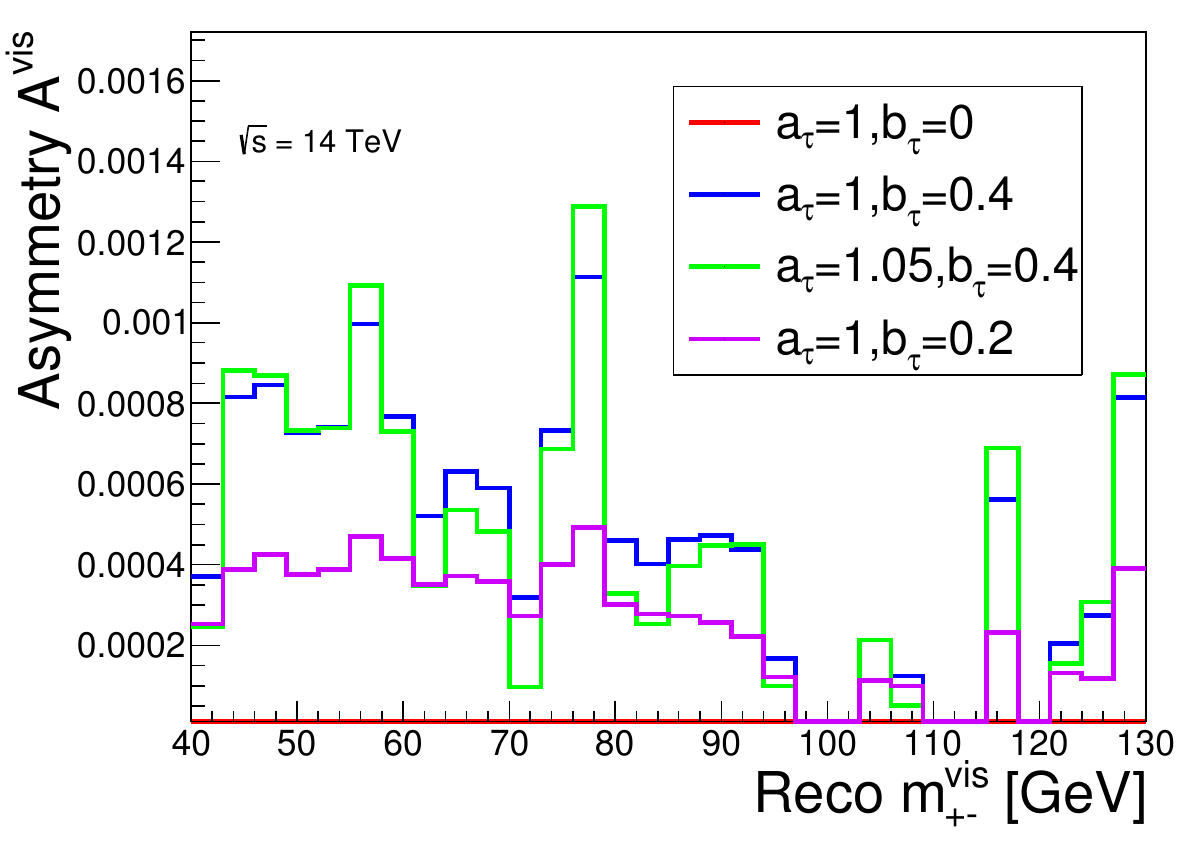}\label{fig:recoplot2}}\\
\subfloat[Visible asymmetry after mass refit
  .]{\includegraphics[width=0.45\linewidth]{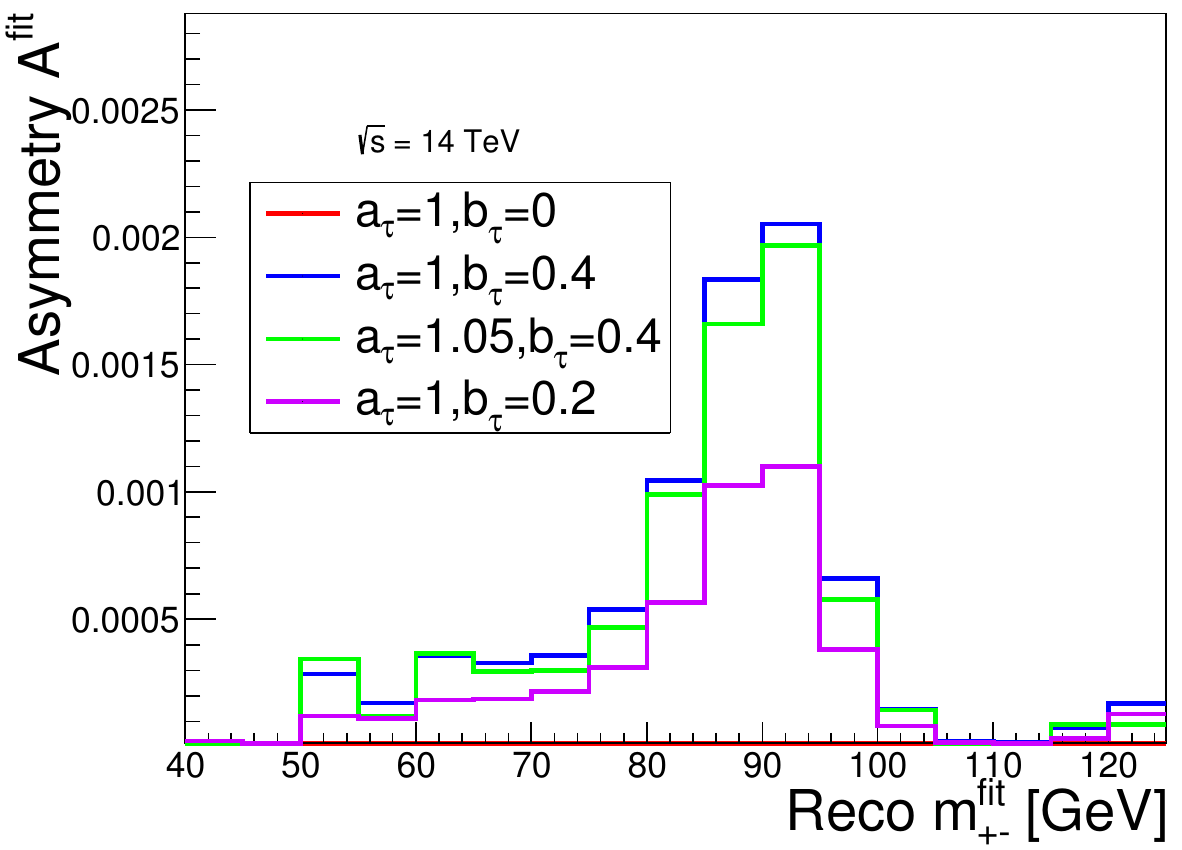}\label{fig:mcplot3_2}} \hspace{5mm}
\subfloat[Data-like asymmetry after mass
  refit.]{\includegraphics[width=0.45\linewidth]{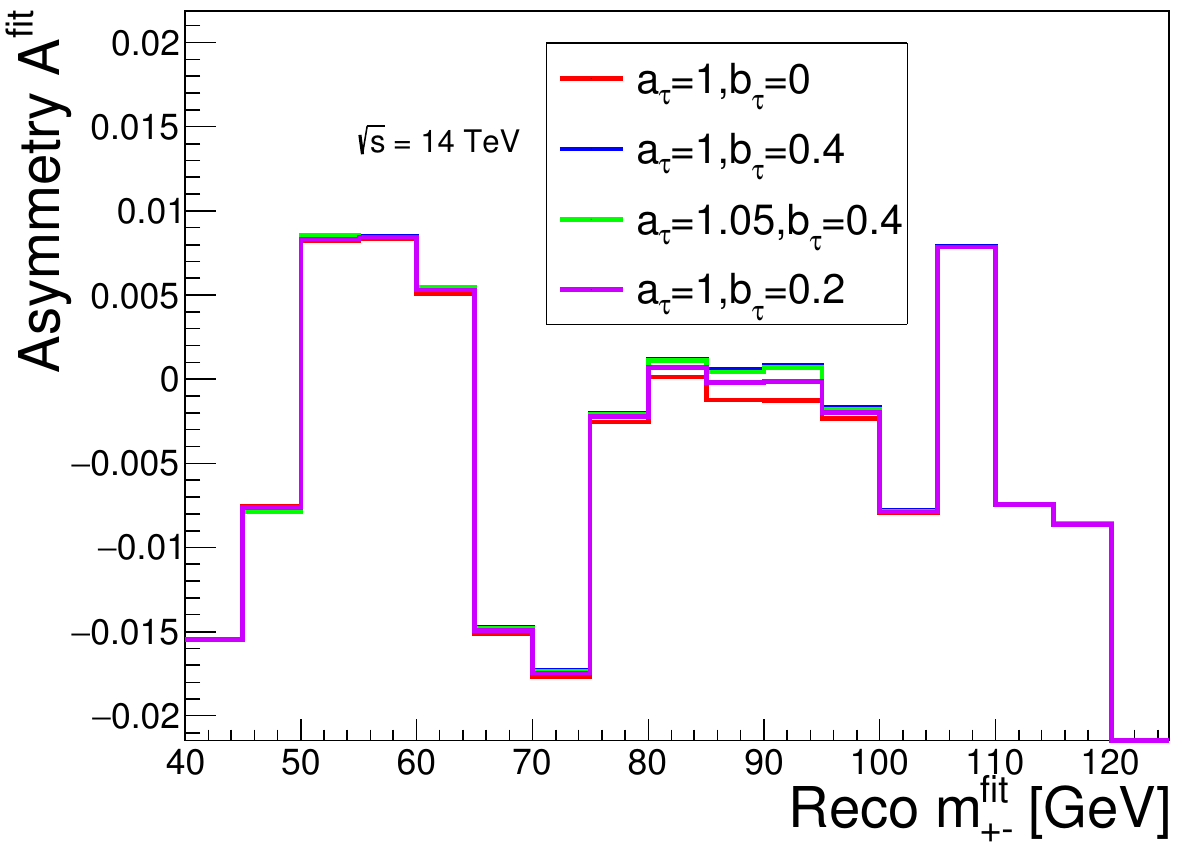}\label{fig:recoplot4}}
\caption{MC-based predictions of the (a)-(b) invariant mass of the
  di-$\tau$ system as well as the (c)-(f) asymmetry sensitivity for
  different values of $a_{\tau}$ and $b_{\tau}$ as a function of
  di-$\tau$ invariant mass.  In (c)-(e) the asymmetry is normalised
  such that it is always $0$ in the SM, while in (f) the statistical
  fluctuation of the MC generator are included, with the number of
  events around 5 times larger than the yields expected at HL-LHC.}
\label{fig:mcplot1}
\end{figure*}

For the calculation of gluon-fusion production of the Higgs boson, the
\texttt{PowhegBox v2} \cite{Frixione:2002ik, Alioli:2010xd,
  Nason:2004rx, Campbell:2012am} generator was used with the
\texttt{NNPDF3.0NNLO} \cite{Anastasiou:2003ds} PDF set.  Proton-proton
collisions are set to happen at center-of-mass energy of $14$ TeV, as
is expected for HL-LHC.  For the simulation of the decay of the Higgs
boson, modelling of the parton showers, and hadronization, the
simulated events were processed with the \texttt{Pythia v8.306}
\cite{Bierlich:2022pfr} program with the \texttt{CTEQ6L1}
\cite{Pumplin:2002vw} PDF set.  \texttt{DELPHES 3.5}
\cite{deFavereau:2013fsa} framework is then used to emulate the
resolution and reconstruction of physical objects (such as photons,
$\tau$ leptons, and jets) by a general-purpose particle detector (such
as ATLAS or CMS) using the "HLLHC" card.  \texttt{FastJet 3.3.4}
\cite{Cacciari:2011ma} package is used to perform the jet clustering
using the anti-$k_t$ algorithm \cite{Cacciari:2008gp}.

In the simulation studies photons are required to have $p_T > 10$~GeV
and to be isolated with an angular cone defined by the
condition\footnote{Here and everywhere we use the cylindrical
coordinates $(r,\phi)$ to describe the transverse plane, $\phi$ being
the azimuthal angle around the beam line.  The pseudorapidity $\eta$
is defined as $-\ln \tan(\theta/2)$.  Finally, the angular distance is
measured in units of $\Delta R \equiv \sqrt{(\Delta\eta)^{2} +
  (\Delta\phi)^{2}}$.} $\Delta R\leq 0.3$.  The reconstructed $\tau$
leptons are required to have $p_T > 15$~GeV.  Their reconstruction is
based on seed jets with the radius parameter~\cite{Cacciari:2008gp}
$\textsf{R} = 0.4$.  This $p_T$ selection represents a realistic lower
limit of what a general purpose detector can achieve.  We assume that
hadronically decaying $\tau$ leptons can be identified with $100\%$
efficiency.  In reality this efficiency will be heavily dependent on
the desired jet rejection power achievable with the conditions of the
HL-LHC.  The results presented in this section scale trivially with
the $\tau$ identification efficiency.  This optimisation is left for
the future, more realistic, simulations of the performance of $\tau$
identification algorithms at the HL-LHC.  All plots in this subsection
are based on the MC simulation described above.

\begin{figure}
\centering
\subfloat[Invariant mass of the
  di-$\tau$]{\includegraphics[width=1.075\linewidth]{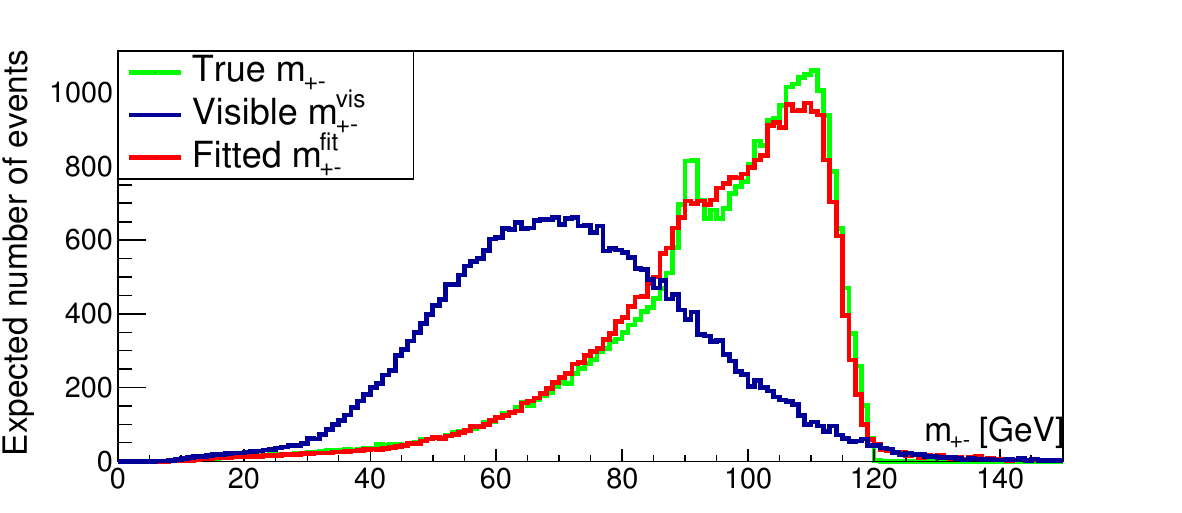}}
\\
\subfloat[Invariant mass of $\tau^{-}
  \gamma$]{\includegraphics[width=1.075\linewidth]{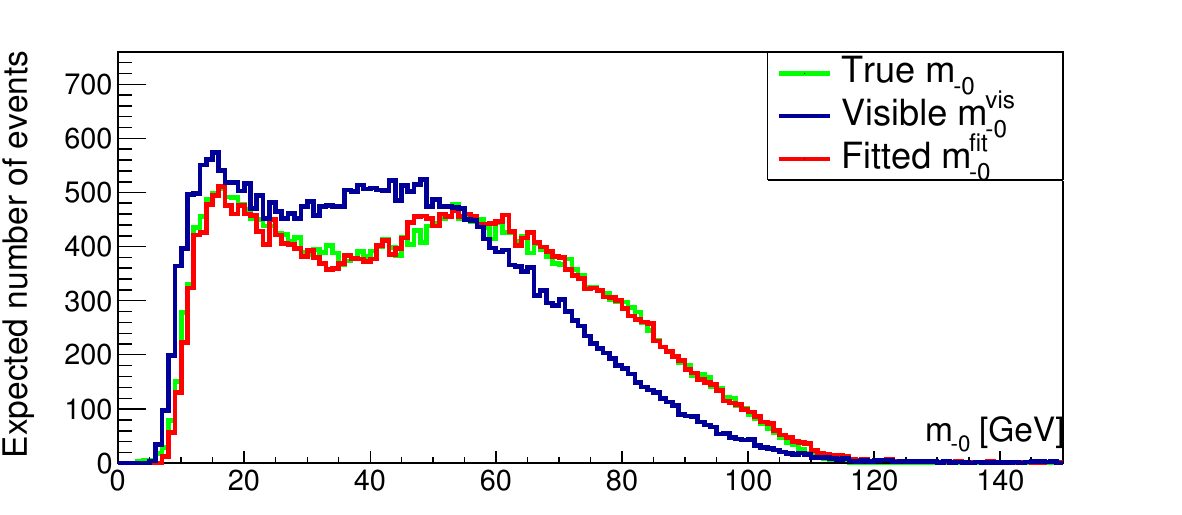}}\\
\subfloat[Invariant mass of $\tau^{+}
  \gamma$]{\includegraphics[width=1.075\linewidth]{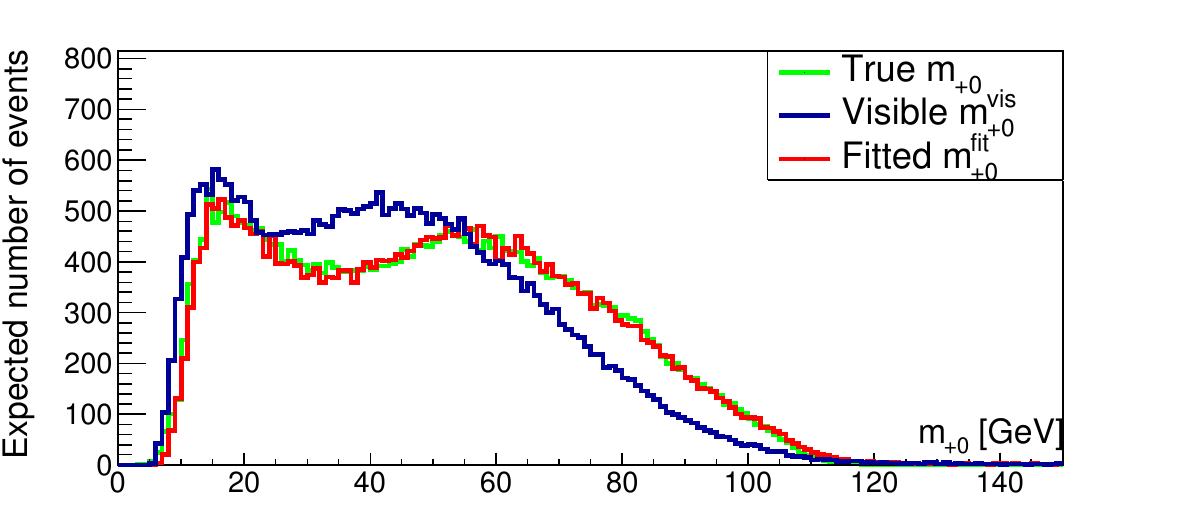}}
\caption{The invariant masses $m_{+-}, m_{+0}$ and $m_{-0}$ using
  true, visible, and fitted physical objects.}
\label{fig:mcplot4}
\end{figure}

The HL-LHC is expected to deliver about $3000$~fb$^{-1}$ integrated
luminosity of data \cite{ZurbanoFernandez:2020cco}.  This corresponds
to over $160$ million events with gluon-gluon fusion production of the
Higgs boson.  With hadronically reconstructed $\tau$s and taking the
same kinematic constraints as considered in Section
\ref{sec:numerical-study}, we estimate that $2.24 \times 10^{5}$ of
these Higgs bosons will eventually decay into the $\gamma \,
\tau^+_\text{had} \, \tau^-_\text{had}$ final state\footnote{This
estimate does not include the laboratory frame requirements on $p_T$,
$\Delta R$ and \textsf{R} discussed above.}.  Approximately $10\%$ of
the events will have the di-$\tau$ system with the invariant mass
$m_{+-}$ within $5$ GeV of the $Z$-boson mass peak where the
forward-backward asymmetry manifests, see Fig.~\ref{fig:massplot1}.
For any selected range of $m_{+-}$ we can estimate the number of
events in `forward' and `backward' regions, say $N_F$ and $N_B$
respectively.  Thus we can easily estimate the following
forward-backward asymmetry,
\begin{equation}
\label{eq:expasym}
A = \frac{N_F - N_B}{N_F + N_B}\, .
\end{equation}

The laboratory frame kinematic requirements applied to the
reconstructed objects, such as the $\tau$ and photon $p_T$ and
isolation requirements, further reduce the number of available events
by a factor of 3 in the $Z$ mass peak region, see
Fig.~\ref{fig:massplot2}.  The photon $p_T$ requirement by itself is
responsible for a $50\%$ decrease in the selection
efficiency.\footnote{This can be contrasted with the fact that the
photon energy cut \textit{in Higgs rest frame} of $5$~GeV has no
effect when $\modulus{m_{+-} - m_Z} \leqslant 5\,\Gamma_Z$, as
mentioned in Sec.~\ref{sec:numerical-study}.}

\subsection{Kinematic Fit}

Although the true invariant mass of the di-$\tau$ system ($m_{+-}$)
offers a good way to access the forward-backward asymmetry, see
Fig.~\ref{fig:recoplot1}, it is not accessible experimentally.  The
short lifetime of the $\tau$ leptons means that they will decay before
reaching the detector, with $\nu_{\tau}$ escaping undetected.  For
hadronically decaying taus that are used in the present study, the
particles registered in the detector will be predominantly charged and
neutral pions.  The detectors have limited acceptance and resolution,
meaning that energies and momenta of these particles will be
reconstructed with a limited accuracy.  The visible invariant mass of
the di-$\tau$ system ($m_{+-}^\textrm{vis}$), constructed from the
visible decay products of $\tau$ decays, offers a degraded sensitivity
to the forward-backward asymmetry, with almost no visible $Z$ peak,
see Fig.~\ref{fig:recoplot2}.  A fit procedure to recover the
sensitivity to the asymmetry based on the kinematic constraints of the
system is described in the following.\footnote{At this point we have
three different ways to compute the invariant masses (and the
asymmetry): true ($m_X, A$), using the full information of the
$\nu_{\tau}$ momentum from the MC; visible ($m_X^\textrm{vis},
A^\textrm{vis}$), using no information about the $\nu_{\tau}$
momentum; and fitted ($m_X^\textrm{fit}, A^\textrm{fit}$), using the
information obtained in the fit procedure.  Here $m_X$ can denote
$m_{+-} = m(\tau^+ \, \tau^-)$, $m_{+0} = m(\tau^+ \, \gamma)$ or
$m_{-0} = m(\tau^- \, \gamma)$.}\\

The final state of $H \to \gamma \, \tau^+ \, \tau^-$ is subject to
two constraints:
\begin{enumerate}
\item The true invariant mass of the three final particles must be
  equal to the mass of the Higgs boson.
\item The energy in the transverse plane, perpendicular to the beam
  line, should be conserved and equal to $0$, with any deviations
  coming from either the missing neutrinos ($\nu_\tau,
  \overline{\nu}_\tau$) or mismeasurements of the particle's energies.
\end{enumerate}  
A fit procedure using \textsf{Minuit2} \cite{James:1975dr} is
performed based on these two conditions with the overall energy of the
two $\tau$ leptons as free parameters.  Since the opening angle
between the neutrinos and visible tau decay product has to be of the
order of $m_\tau / E_\tau$, both $\nu_{\tau}$ and $\bar{\nu}_\tau$ are
predominantly collinear with the visible parts of the hadronically
decaying $\tau$'s, for the energies considered here.  Therefore, the
approach of treating the contributions from the $\tau$ neutrino and
$\tau$ energy smearing as one common parameter that only affects the
energy of the $\tau$-lepton and not its spacial direction is
justified.

This simple fit procedure allows us to restore the true energies of
the $\tau$-leptons and the fitted two-body invariant masses match well
with the true invariant masses, as demonstrated in
Fig.~\ref{fig:mcplot4}.  Further the fitted invariant masses
$m^\text{fit}_{+0}$ and $m^\text{fit}_{-0}$ are used to identify
events in forward and backward regions.  This information is then used
to estimate the asymmetry $A^\text{fit}$ while selecting
$m^\text{fit}_{+-}$ in the region around $Z$-boson mass where the
asymmetry is maximal.  The asymmetry defined using the fitted masses,
$A^\text{fit}$ behaves similarly as expected for the true asymmetry as
a function of the fitted di-$\tau$ mass, see Fig.~\ref{fig:mcplot3_2}.
Thus $A^\text{fit}$ is a reasonable estimator of the forward-backward
asymmetry.  In the following we evaluate this asymmetry in a
real-data-like environment.

\subsection{Asymmetry Calculation}

\begin{figure*}[hbtp]
\centering
\subfloat[Truth-based asymmetry
  fits.\label{fig:fitplot1a}]{\includegraphics[width=0.33\linewidth]{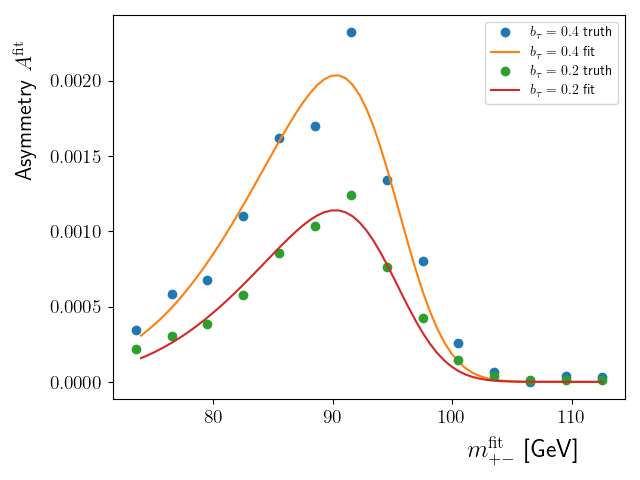}}
\hfill
\subfloat[\raisebox{-10pt}{\parbox{0.85\linewidth}{Fits to the reconstructed asymmetry distribution 
  at $b_{\tau} = 0.1$, with $b_{\tau} = 0.4$ fit
  overlaid.}}\label{fig:fitplot1b}]{\includegraphics[width=0.33\linewidth]{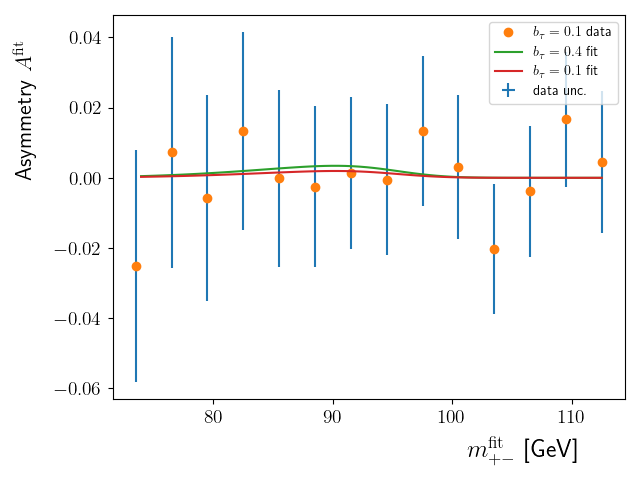}} \hfill
\subfloat[$b_{\tau}$ vs scaling factor
  $c$.\label{fig:fitplot1c}]{\includegraphics[width=0.33\linewidth]{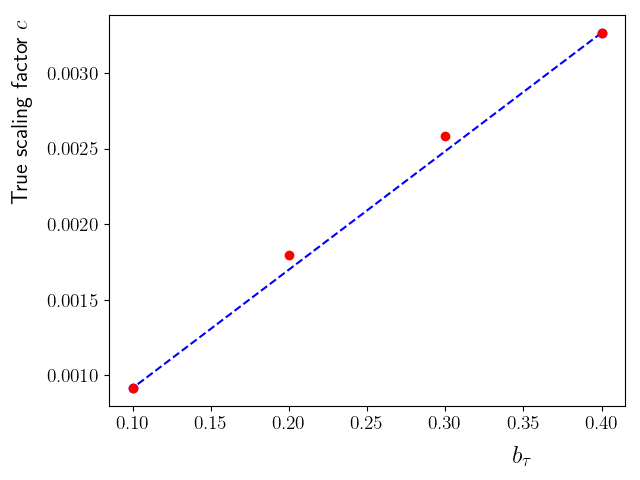}}
	\caption{Gaussian fits to (a) truth-based asymmetry shapes and (b)
data-like shapes in MC with $b_{\tau}=0.1$ with the $b_{\tau}=0.4$ fit overlaid.
 (c) shows the relationship between true $b_{\tau}$ and the scaling factor $c$ from
  the fits of analytically computed asymmetry $A$ distributions.
  Note that the uncertainty bars in (b) represent the
  statistical uncertainty expected at HL-LHC.}
\label{fig:fitplot1}
\end{figure*}

We compare two methods of quantifying the asymmetry and estimating the
corresponding values of $b_{\tau}$.  The first approach uses a simple
selection of events with $m^\text{fit}_{+-}$ close to the $Z$ mass
peak, where the asymmetry is maximised.  Here, the window of $\pm 9$
GeV around $m_Z$ was chosen, i.e.\ $\modulus{m^\text{fit}_{+-} - m_Z}
\leqslant 9$~GeV.  The width of this window was inspired by the range
of the di-tau mass $m^\text{fit}_{+-}$ where the asymmetry is
enhanced, see Figs.~\ref{fig:mcplot3_2} and \ref{fig:recoplot4}.  The
asymmetry estimate $A^\text{fit}$ is then computed as in the equation
\ref{eq:expasym} and used to predict $b_\tau$.

The second approach involves widening the di-$\tau$ mass selection to
the range of $72$--$114$ GeV.  The asymmetry $A^\text{fit}$ is
computed in bins of $3$ GeV.  A skewed Gaussian $f_\text{skew}(t)$ is
then fitted to the shape of the asymmetry distribution,
\begin{equation}
	f_\text{skew}(t) = c \phi(t) \Phi(\alpha t), \qquad t = \frac{m_{+-}^\text{fit}-a}{b},
\end{equation}
where $\phi(t)$ is the normal probability density function, $\Phi(t)$
is the normal cumulative distribution function, and $a,b,c,\alpha$ are
the free parameters in the fit.  
The parameters can be determined by fitting the true asymmetry distributions for specific values of $b_\tau$ as shown in Fig.~\ref{fig:fitplot1a}. Except for the parameter $c$, which determines the height of the skewed Gaussian and depends on $b_\tau$, all other parameters agree for different $b_\tau$ values (within statistical uncertainty). Keeping all parameters except the overall scaling factor $c$ fixed at the determined values, we fit the distribution of asymmetry $A^\text{fit}$ in the reconstructed MC, which gives us the fitted value of $c$. As an example the dataset corresponding to $b_\tau = 0.1$ is shown in Fig.~\ref{fig:fitplot1b} along with the fitted skewed Gaussian distribution. For comparison, the skewed Gaussian corresponding to the case of $b_\tau = 0.4$ fit is overlaid. In Fig.~\ref{fig:fitplot1c} we show that the scaling factor $c$ obtained from fitting is directly proportional to $b_\tau$. Thus knowing $c$ one can directly infer the value of $b_\tau$.

\begin{table}[htbp]
\centering
\begin{tabular}{c|c|c}
\toprule
$(a_{\tau},b_{\tau})$ & $b_{\tau}$ from $\pm 9$ GeV  &
$b_{\tau}$ from  \\
& mass window & $f_\text{skew}(x) $ fit \\
\midrule
$(1, 0.0)$ & $0.22 \pm 2.24$ & $0.23 \pm 2.06$ \\
$(1,0.1)$ & $ 0.32 \pm 2.24$ & $0.32 \pm 2.07$\\
$(1,0.2)$ & $ 0.41 \pm 2.22$ & $0.44 \pm 2.14$\\
$(1,0.3)$ & $ 0.53 \pm 2.36$ & $0.55 \pm 2.24$\\
$(1,0.4)$ & $ 0.65 \pm 2.52$ & $0.67 \pm 2.39$\\
\bottomrule 
\end{tabular}
\caption{True and predicted values of $b_{\tau}$ from two different
  methods.  Integrated luminosity of $3000$fb$^{-1}$ is assumed.}
\label{tab:preds}
\end{table}

The results of the two approaches are summarised in Table
\ref{tab:preds}.  The uncertainties of the measurements include the
statistical uncertainty of the expected HL-LHC event yields, which is
the dominant one.  To estimate the HL-LHC uncertainty contribution we
rescale the yields to match those expected at $3000$fb$^{-1}$ and
recompute the statistical uncertainty accordingly.  Both of the
approaches produce comparable central values with the fit to
$f_\text{skew}$ resulting in lower uncertainties. Note that an offset between the theoretical input values and predicted central values of $b_\tau$ is present, due to a statistical fluctuation in the MC sample. The offset remains constant for all tested $b_\tau$ values and is reproduced by both methods tested, further validating their stability.

\subsection{Backgrounds Contributions and its Separation} 

\begin{figure*}[hbtp]
\centering
\subfloat[$\Delta R$ between leading tau and $\gamma$
]{\includegraphics[width=0.45\linewidth]{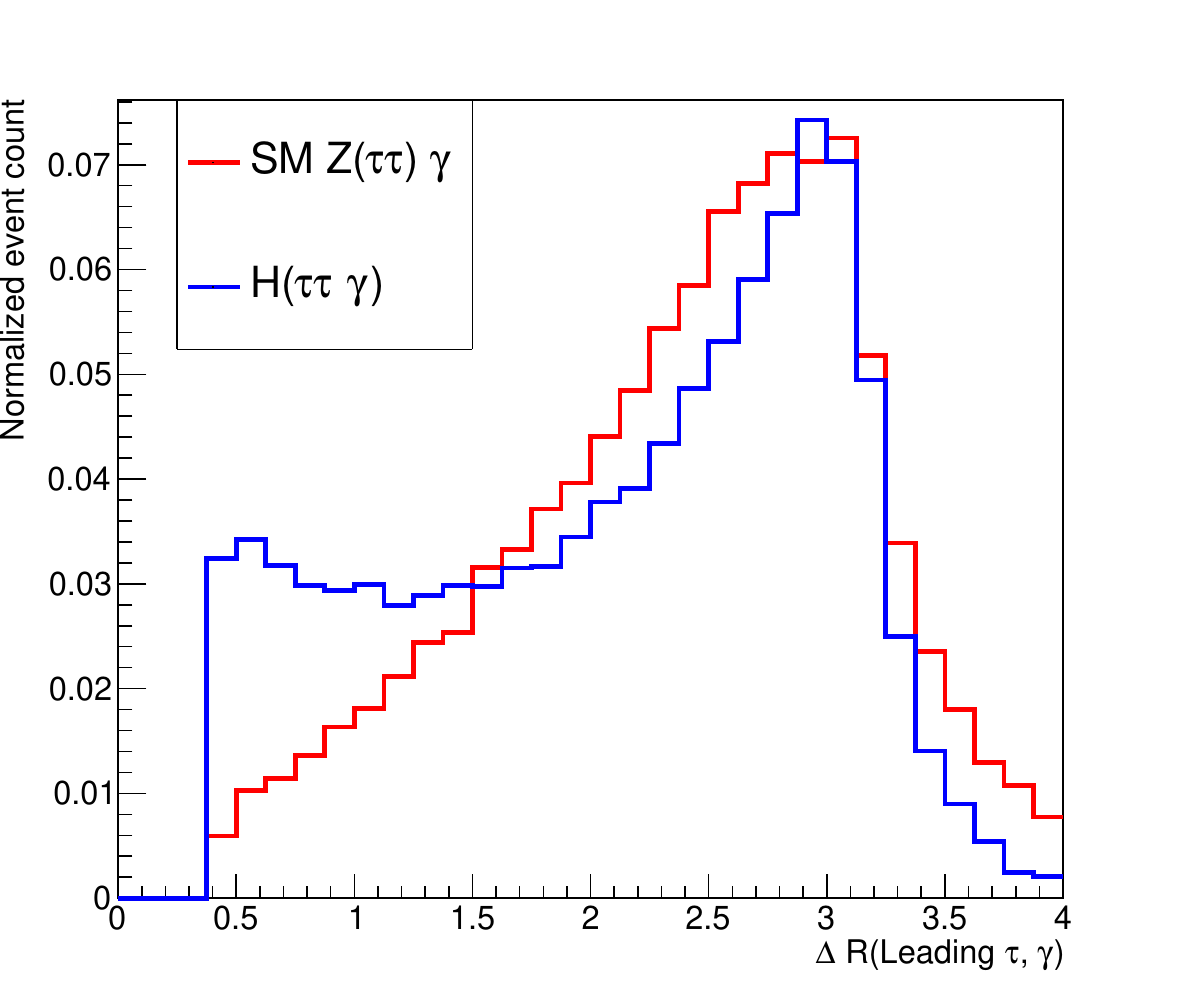}\label{fig:bkgplot1}}
\subfloat[$\Delta R$ between sub-leading tau and $\gamma$
]{\includegraphics[width=0.45\linewidth]{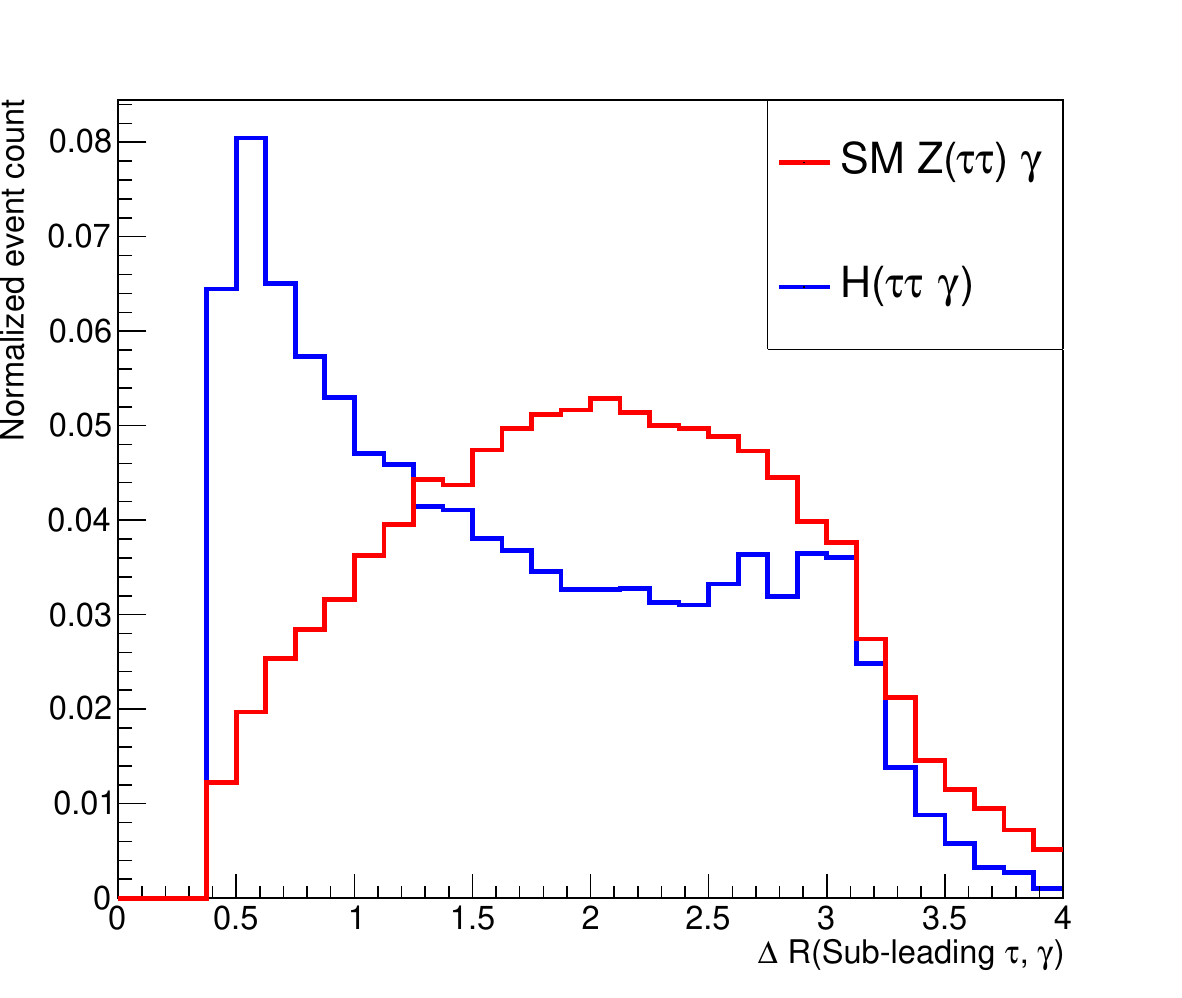}\label{fig:bkgplot2}}\\
\subfloat[$\Delta R$ between di-tau and $\gamma$
]{\includegraphics[width=0.45\linewidth]{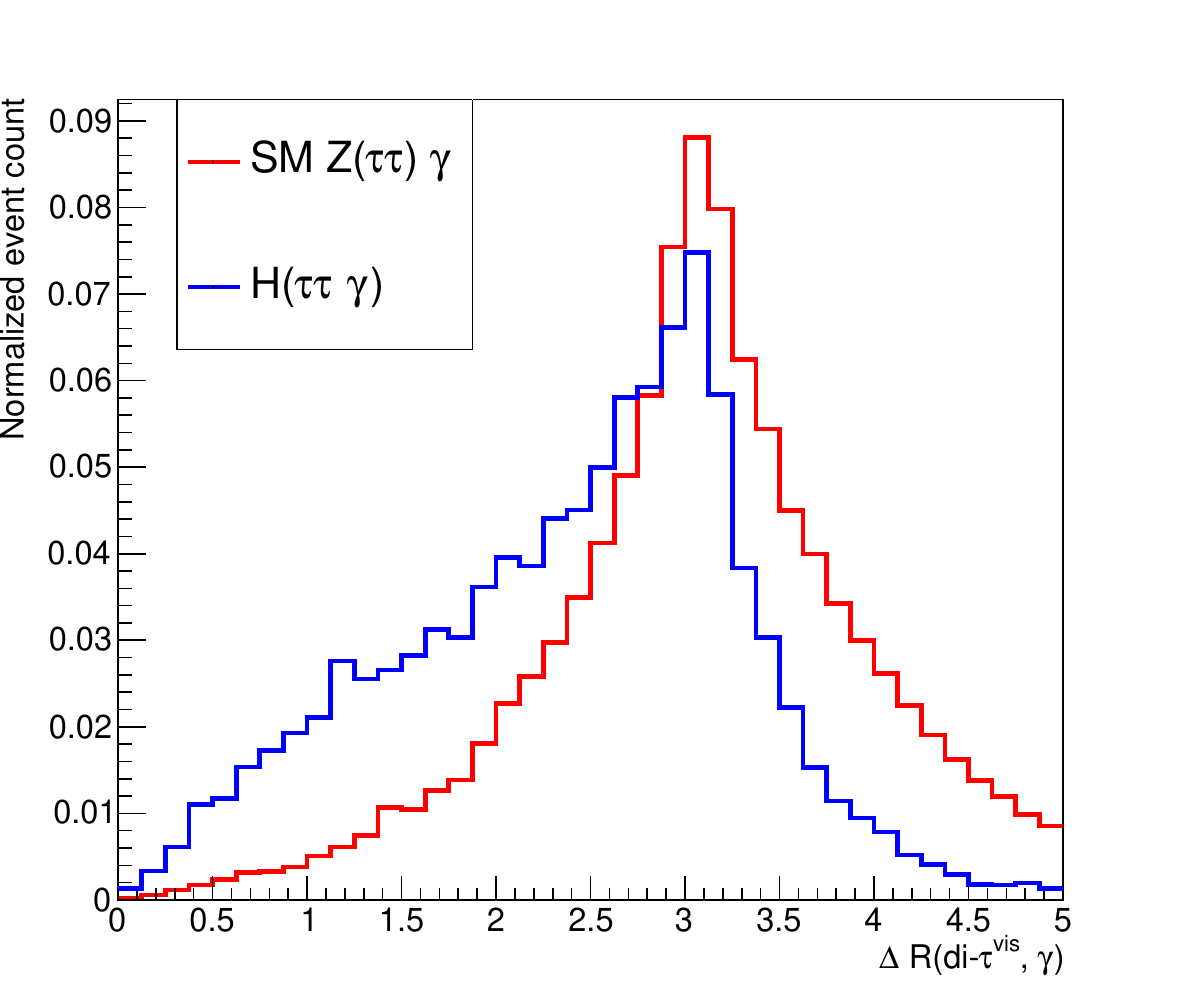}\label{fig:bkgplot3}}
\subfloat[Leading $\gamma$ $p_T$
]{\includegraphics[width=0.45\linewidth]{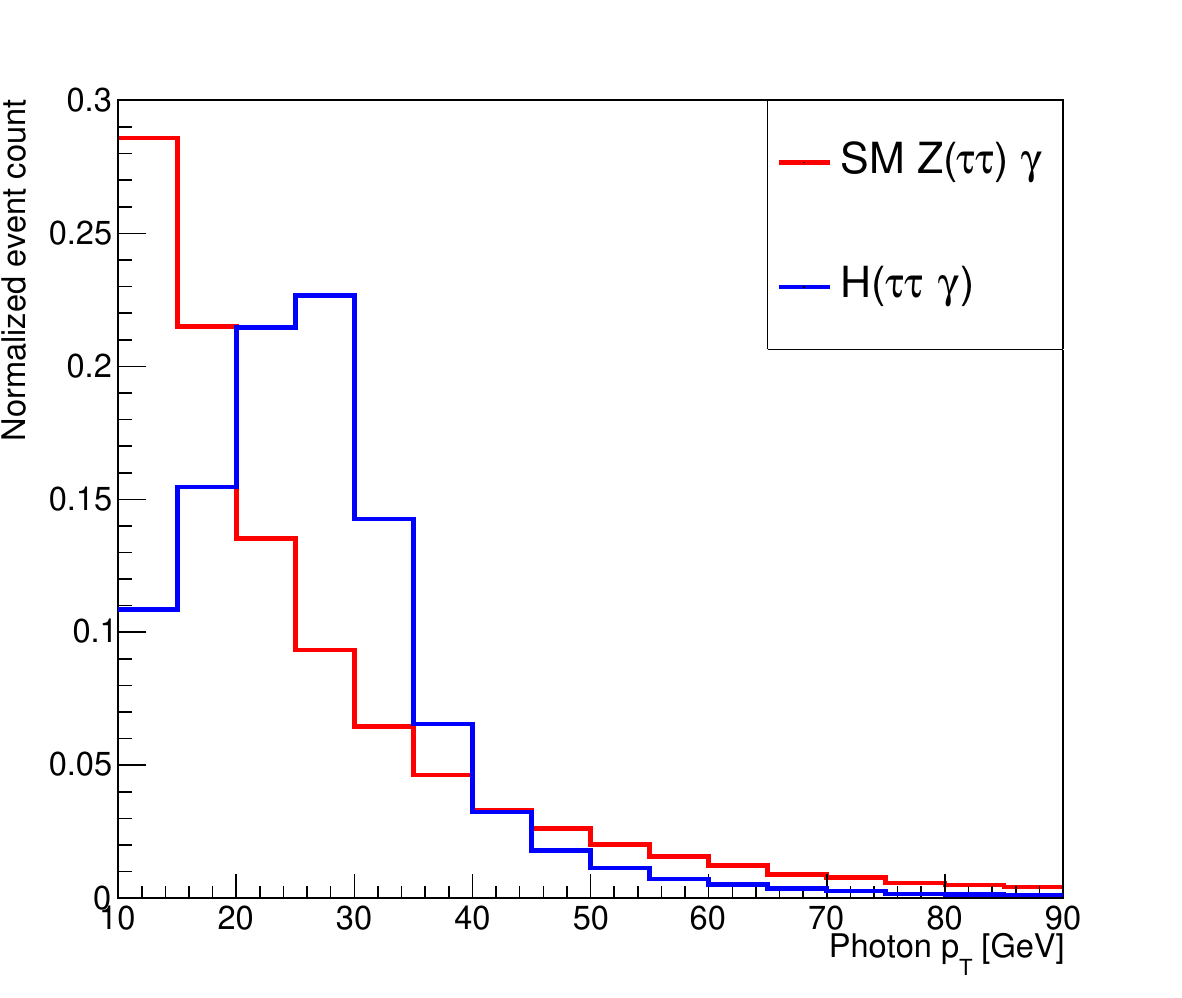}\label{fig:bkgplot4}}\\

\caption{MC-based comparisons of the $H\to\tau\tau\gamma$ signal and $Z+\gamma$ background samples, showing (a)-(b) the angular distance between the $\tau$-leptons and photon, (c) angular distance between the di-tau system and photon, (d) the transverse momentum of the leading photon. Both samples are normalized to 1 for ease of comparison.}
\label{fig:bkgplot}
\end{figure*}

The dominant SM background we have to consider is $Z(\tau\tau)+\gamma$ production, similar to the ATLAS \cite{ATLAS:2020qcv} and CMS \cite{CMS:2022ahq}  $H \to Z(\ell\ell) \gamma$ searches. However, compared to the $\ell\ell$ channel, in the $\tau\tau$ channel the non-resonant Higgs decay contribution is dominating, resulting in the much larger (Higgs)/(non-Higgs backgrounds) fraction in the $Z$ mass peak region of $m_{+-}$. The kinematics of the events also change, allowing for easier discrimination of the background. Developing an algorithm for suppression of the $Z+\gamma$ backgrounds (which most likely would have to be done with the application of machine learning techniques to fully utilize several correlated kinematic values) is beyond the scope of this paper. However to illustrate the possibility of such classifiers we perform a simple MC comparison.

The generation of $Z+\gamma$ MC is done using \texttt{MadGraph5 aMC@NLO 3.5} \cite{Alwall:2014hca} with the \texttt{NNPDF3.0NLO} \cite{Anastasiou:2003ds} PDF set. Further decay chains and hadronization are handled by \texttt{Pythia} and \texttt{DELPHES} with the same setup as described in Section \ref{sec:mc}. We note several key kinematic variables related to $\tau$-leptons and photons that can help discriminate between signal and background, such as $\Delta R(\tau,\gamma)$, $\Delta R(Z^\text{vis},\gamma)$, photon $p_T$, see Figure \ref{fig:bkgplot}. Higgs candidate $p_T$ and invariant mass can also be useful, as has been shown in the ATLAS and CMS searches. Based on this we believe that an efficient classifier can be built to significantly suppress the background contribution. The uncertainties quoted in Table~\ref{tab:preds} will increase in the presense of background events, but the exact effect depends strongly on how well the signal/background separation can be performed.


\section{Conclusions}\label{sec:conclusion}

We have analysed the 3-body decay of the Higgs boson $H \to \tau^+ \,
\tau^- \, \gamma$ as an additional source of information about the CP
violation in the $H\tau\tau$ Yukawa coupling, independent from the
existing experimental studies on the 2-body decay $H \to
\tau^+\,\tau^-$ \cite{ATLAS:2020evk, CMS:2021sdq, CMS:2022uox,
  Berge:2008wi, Berge:2008dr, Berge:2011ij, Berge:2013jra,
  Harnik:2013aja, Hagiwara:2016zqz}.  The forward-backward asymmetry
in the $\tau$ angular distribution in our case arises due to the
interference of the tree-level contribution (which includes the CP
violating $H\tau\tau$ Yukawa coupling $b_\tau \neq 0$) and the CP-even
SM loop-level contributions.  We have proposed a novel method of
measuring forward-backward asymmetry in the Dalitz plot distribution
of events in the plane of $\gamma\,\tau^\pm$ Lorentz invariant masses
($m_{+0}$ vs.\ $m_{-0}$ plane).  Such a Dalitz plot distribution is
frame independent, making the method of extraction of forward-backward
asymmetry clean and attractive from the experimental point of view.
The asymmetry is directly proportional to the CP-odd $H\tau\tau$
coupling parameter $b_\tau$.  In principle, the asymmetry can also
appear from the interference of CP-even tree-level contribution and CP
violating loop-level contributions.  However, for our numerical study,
we assume no CP-violation at loop-level and focus only on the effects
of non-zero $b_\tau$ and whether this can be experimentally probed at
HL-LHC.

The forward-backward asymmetry is predicted to be the largest when the
di-$\tau$ invariant mass $m_{+-}$ is close to $m_Z$ (it could reach
$\sim 1\%$ for high values of $b_\tau$) and it rapidly diminishes as
one moves farther away from the $Z$ pole.  To estimate the feasibility
of such asymmetry measurements at the HL-LHC we have performed a
simplified MC simulation with kinematic cuts meant to mimic the
experimental conditions.  A kinematic fit was used to constrain the
hadronically reconstructed $\tau$-leptons and account for the missing
$\nu_{\tau}$ information not available in the detector.  We estimated
the asymmetry directly in the region with di-$\tau$ mass in the range
of $m_Z \pm 9$~GeV for different values of $b_\tau$.  We also looked
for the asymmetry by performing a shape fit in a wider mass region,
$72~\text{GeV} \leqslant m_{+-} \leqslant 114~\text{GeV}$.

From our MC studies we find that the statistical uncertainties we
currently expect to get with the HL-LHC dataset are significantly
larger than the effect itself.  Nevertheless, our simplistic MC study
suggests that our proposed methodology is experimentally doable, and
our results could be encouraging for more detailed and in-depth
explorations in the future.  Instead of the one-dimensional binned
shape fit used in this study a full two-dimensional unbinned Dalitz
plot analysis could instead be envisaged using for example the Miranda
method \cite{Bediaga:2009tr,BaBar:2008xzl}, the method of energy test
statistic \cite{aslan2005new, Williams:2011cd, LHCb:2014nnj,
  LHCb:2023mwc, LHCb:2023rae} and the earth mover's
distance~\cite{Davis:2023lxq}.  The asymmetry can also appear from the
interference of CP-even tree-level contribution and CP violating
loop-level contributions, this effect has not been considered in our
numerical studies yet.  In our MC simulation, we have only considered
final states with both of the $\tau$-leptons decaying hadronically,
the dataset can be doubled by also considering one of the $\tau$s to
decay leptonically, i.e.\ adding the $H \to \tau_\text{had}
\tau_\text{lep} \gamma$ decay channel.  With the better understanding
of the technical capabilities of particle detectors such as ATLAS and
CMS after the Phase-2 upgrades, the kinematic selections can be
further optimised.

Finally, once the asymmetry can be probed with reduced uncertainty, it
would be interesting to compare its prediction for $b_\tau$ with that
obtained from the already ongoing experimental study of $H \to \tau^+
\, \tau^- \to m^+ \, \overline{\nu}_\tau \, m^- \, \nu_\tau$ where
$m=\pi,\rho$ etc.  If there is significant deviation in the two
$b_\tau$ values, one can assume that there is some significant
CP-violation coming from the loop-level contribution, which we have
neglected in our numerical study in this paper.  It is interesting to
note that the same loop-level diagrams also contribute to $H \to
\ell^+ \, \ell^- \, \gamma$ for $\ell=e,\mu$, and for these decay
modes the tree-level contributions are negligible.  Moreover, the same
Dalitz plot techniques developed for $H \to \tau^+ \, \tau^- \,
\gamma$ can also be applied to probe the asymmetry in the Dalitz plots
of $H \to \ell^+ \, \ell^- \, \gamma$ to constrain or discover the CP
violation at loop-level.  Therefore, our formalism of probing the
forward-backward asymmetry inside the Lorentz invariant Dalitz plot
distribution of events would certainly help explore CP property of the
Higgs boson in a more systematic and unified manner.

\section*{Acknowledgements} 
We thank Steffen M{\ae}land and Bjarne Stugu for helpful discussions
concerning experimental signatures of the CP violation in the
$H\tau\tau$ Yukawa coupling. This research has received funding from
the Norwegian Financial Mechanism for years 2014-2021, under the grant
no 2019/34/H/ST2/00707.  The work of DS is supported by the Polish
National Science Centre under the Grant number
DEC-2019/35/B/ST2/02008.

\bibliographystyle{utphys_jr}
\bibliography{References}

\end{document}